\documentclass[useAMS,usenatbib]{mn2e}
\usepackage{graphicx}

\def\I{\'\i}

\title[16 (EN) Lac and 2 And]{The 2003-4 multisite photometric campaign for the $\bbeta$~Cephei and eclipsing 
star 16 (EN) Lacertae with an Appendix on 2 Andromedae, the variable comparison star}
\author[M. Jerzykiewicz et al.]{M.~Jerzykiewicz,$^{1}$\thanks{E-mail: jerzykiewicz@astro.uni.wroc.pl} 
G.~Handler,$^{2}$ J.~Daszy{\'n}ska-Daszkiewicz,$^{1}$ A.~Pigulski,$^{1}$\newauthor E.~Poretti,$^{3}$ 
E.~Rodr\I guez,$^{4}$ P.~J.~Amado,$^{4}$ Z.~Ko{\l}aczkowski,$^{1}$ K.~Uytterhoeven,$^{5, 6}$\newauthor 
T.~N.~Dorokhova,$^{7}$~N.~I.~Dorokhov,$^{7}$ D.~Lorenz,$^{8}$ D.~Zsuffa,$^{9}$ S.-L.~Kim,$^{10}$\newauthor 
P.-O.~Bourge,$^{11}$ B.~Acke,$^{5}$ J.~De Ridder,$^{5}$ T.~Verhoelst,$^{5}$  R.~Drummond,$^{5}$\newauthor
A.~I.~Movchan,$^{7}$ J.-A.~Lee,$^{10}$ M.~St\c{e}\'slicki,$^{12}$ J.~Molenda-\. Zakowicz,$^{1}$ 
R.~Garrido,$^{4}$\newauthor S.-H.~Kim,$^{10}$ G.~Michalska,$^{1}$ M.~Papar\'o,$^{9}$ V.~Antoci,$^{8}$ 
C.~Aerts$^{5}$ 
\and \\  
$^{1}$ Astronomical Institute of the Wroc{\l}aw Univeristy, Kopernika 11, 51-622 Wroc{\l}aw, Poland\\ 
$^{2}$ Copernicus Astronomical Center, Bartycka 18, 00-716, Warsaw, Poland\\ 
$^{3}$ INAF-Osservatorio Astronomico di Brera, Via Bianchi 46, 23807 Merate, Italy\\
$^{4}$ Instituto de Astrofisica de Andalucia, C.S.I.C., Apdo. 3004, 18080 Granada, Spain\\
$^{5}$ Instituut voor Sterrenkunde, K. U. Leuven, Celestijnenlaan 200B, B-3001 Leuven, Belgium\\
$^{6}$ Mercator Telescope, Calle Alvarez de Abreu 70, 38700 Santa Cruz de La Palma, Spain\\
$^{7}$ Astronomical Observatory of Odessa National University, Marazlievskaya, 1v, 65014 Odessa, Ukraine\\
$^{8}$ Institut f\"ur Astronomie, Universit\"at Wien, T\"urkenschanzstrasse 17, A-1180 Wien, Austria\\
$^{9}$ Konkoly Observatory, MTA CSFK, Konkoly Thege Mikl\'os \'ut 15-17., 1121 Budapest, Hungary\\
$^{10}$ Korea Astronomy and Space Science Institute, Daejeon, 305-348, Korea\\
$^{11}$ Institut d'Astrophysique et de G\'eophysique, Universit\'e de Li\`ege, all\'ee du Six Ao\^{u}t 17, 
4000 Li\`ege, Belgium\\
$^{12}$ Space Research Centre, Polish Academy of Sciences, Solar Physics Division, ul. Kopernika 11, 51-622 
Wroc{\l}aw, Poland}

\input epsf
\usepackage{graphics}
\begin{document}
\date{Accepted ... Received ...; in original form ...}
\maketitle
\begin{abstract} 
A multisite photometric campaign for the $\beta$ Cephei and eclipsing variable 16 Lacertae is reported. 749 h of high-quality 
differential photoelectric Str\"omgren, Johnson and Geneva time-series photometry were obtained with ten telescopes during 185 
nights. After removing the pulsation contribution, an attempt was made to solve the resulting eclipse light-curve by means of the 
computer program {\sc EBOP}. Although a unique solution was not obtained, the range of solutions could be constrained by 
comparing computed positions of the secondary component in the Hertzsprung-Russell diagram with evolutionary tracks. 

For three high-amplitude pulsation modes, the $uvy$ and the Geneva $UBG$ amplitude ratios are derived and compared with the 
theoretical ones for spherical-harmonic degrees $\ell \leq{}$4. The highest degree, $\ell ={}$4, is shown to be incompatible with 
the observations. One mode is found to be radial, one is $\ell ={}$1, while in the remaining case $\ell ={}$2 or 3. 

The present multisite observations are combined with the archival photometry in order to investigate the long-term variation of the 
amplitudes and phases of the three high-amplitude pulsation modes. The radial mode shows a non-sinusoidal variation on a time-scale 
of 73 yr. The $\ell ={}$1 mode is a triplet with unequal frequency spacing, giving rise to two beat-periods, 720.7 d and 29.1 yr. 
The amplitude and phase of the $\ell ={}$2 or 3 mode vary on time-scales of 380.5 d and 43 yr. 

The light variation of 2 And, one of the comparison stars, is discussed in the Appendix. 
\end{abstract}

\section{Introduction}

16 (EN) Lacertae $=$ HR\,8725 (B2\,IV, $V ={}$5.59), a member of Lac OB1a, is a single-lined spectroscopic binary \citep{L10, SB25} 
and an eclipsing variable. The orbital period, derived from the epochs of minimum light, $P_{\rm orb} ={}$12.09684 d \citep{J80, 
PJ88}. The system consists of a well-known $\beta$~Cephei variable and an undetected secondary. The $\beta$ Cephei variation of the 
primary is dominated by three pulsation modes with frequencies close to 6 d$^{-1}$. According to \citet[][henceforth F69]{F69}, who 
based his analysis on archival 1951, 1952 and 1954 radial-velocity and photoelectric blue-light observations, the first two modes 
have constant amplitudes, while the amplitude of the third mode varies on a time scale of months. F69 described the first two modes 
as singlets, having frequencies $f_1 ={}$5.91134 and $f_2 ={}$5.85286 d$^{-1}$, and the third mode as a doublet, consisting of two 
terms with frequencies $f_{3,1} ={}$5.49990 and $f_{3,2} = f_1 - 5f_{\rm orb} ={}$5.49799 d$^{-1}$.  Comparable frequencies were 
derived from the 1964 and 1965 photoelectric observations, obtained by one of us (MJ) at Lowell Observatory, viz. $f_1 
={}$5.91120$\,\pm\,$0.00005, $f_2 ={}$5.85503$\,\pm\,$0.00010, and $f_3 ={}$5.50322$\,\pm\,$0.00009 d$^{-1}$ \citep{JJ+79}. The 
first value agrees with that of F69, but the other two differ from F69's $f_2$ and $f_{3,1}$ by 0.0022 and 0.0033 d$^{-1}$, 
respectively. From these differences, \citet{JJ+79} concluded that F69's values of $f_2$ and $f_{3,1}$ suffered from an error of one 
cycle per year (yr$^{-1}$). 

An analysis of all photometric observations of 16 Lac obtained throughout 1992 was carried out by \citet[][henceforth JP96, 
JP99]{JP96, JP99}. The main results of the analysis can be summarized as follows: (1) the amplitudes of the $f_1$ and $f_2$ modes 
vary on a time-scale of decades, the reciprocal time-scales amounting to 0.014 and 0.020 yr$^{-1}$, respectively, (2) the third mode 
is confirmed to be a doublet, but with frequencies different from those derived by F69, viz.\ $f_{3,1} 
={}$5.5025779$\,\pm\,$0.0000005 and $f_{3,2} ={}$5.5040531$\,\pm\,$0.0000008 d$^{-1}$. Note that none of these frequencies bears a 
simple numerical relation to the orbital period. Moreover, JP96 demonstrated that there is no correlation of the pulsation 
amplitudes with the orbital phase. 

The radial-velocity (RV) data available at the time were shown by JP96 to be consistent with the above-mentioned photometric 
results. The RV data, however, were much less numerous than the photometric data, making this conclusion somewhat uncertain. The 
situation has improved after \citet[][henceforth L01]{L+01}, provided new RV observations, more than doubling the number of existing 
measurements. A periodogram analysis of these and the older data led L01 to a number of frequency solutions, the details of which 
depended on the weights assigned to the RV measurements from a particular source. As far as the periods are concerned, the results 
valid for all weighting schemes can be summarized as follows: (1) the first mode is an equidistant triplet with the central term 
having the largest amplitude and a period, $P_1$, equal to 0.16916707~d ($f_1 ={}$5.9113160 d$^{-1}$). The remaining two periods,  
$P^+_1$ and $P^-_1$ in the notation of L01, are equal to 0.16916605 and 0.16916809 d, respectively. The frequency separation of the 
triplet amounts to 0.0000356 d$^{-1}$, corresponding to the amplitude-modulation frequency of 0.0130 yr$^{-1}$, (2) the second mode 
is a doublet; in the order of decreasing amplitude, the periods, $P_2$ and $P^+_2$ in the notation of L01, are equal to 0.17085553 
and 0.17077074 d, respectively ($f_2 ={}$5.8528981 and $f^+_2 ={}$5.8558041 d$^{-1}$), resulting in a beat period of 344 d (the beat 
frequency of 0.00291 d$^{-1}$), (3) the third mode is also a doublet; in the order of decreasing amplitude the periods are $P_3 
={}$0.18173251 and $P^+_3 ={}$0.18168352 d ($f_3 ={}$5.5025928 and $f^+_3 ={}$5.5040765 d$^{-1}$); in this case the beat period is 
equal to 674 d. Conclusions (1) and (3) approximately agree with the results of JP96 and JP99, but conclusion (2) does not. The 
disagreement is twofold. First, the time-scale of the amplitude modulation of the $f_2$ term derived by L01 is less than one 
fiftieth of that derived by JP96 and JP99. Second, L01's $f_2$, i.e.\ the frequency of the higher-amplitude component of the $f_2$, 
$f^+_2$ doublet has a value close to that originally determined by F69 and dismissed by \citet{JJ+79} as a yearly alias of the 
photometric $f_2$ value they derived. Note that $f_2$ of \citet{JJ+79} is close to $f^+_2$, the frequency of the smaller-amplitude 
component of the $f_2$, $f^+_2$ doublet. This, of course, is the consequence of the doublet's beat-period having its value close to 
1 yr. 

In addition to the three modes just discussed, six fainter terms were detected by \citet{J93} in his 1965 data. The $y$ amplitudes 
of these terms ranged from 2.1$\,\pm\,$0.14 to 0.7$\,\pm\,$0.14 mmag, and the frequencies were equal to (in the order of decreasing 
amplitude) 0.1653, 7.194, 11.822, 11.358, 11.414, and 11.766 d$^{-1}$. The first of these is equal to twice the orbital frequency, 
suggesting ellipsoidal variability. However, the observed amplitude and phase excluded this possibility. The third frequency is 
equal to 2$f_1$ and the three last frequencies are the combination terms $f_2 + f_3$, $f_1 + f_3$ and $f_1 + f_2$. The 
7.194-d$^{-1}$ term was attributed by \citet{J93} to an independent pulsation mode and was used as such, together with the three 
strongest ones, in an asteroseismic study of the star \citep{DJ96}. Subsequently, however, this term was shown by \citet{S+97} and 
\citet{HJR06} to be due to a light variation of 2 Andromedae, used by \citet{J93} as a comparison star. 

During a multisite photometric campaign carried out between 2 August 2003 and 9 January 2004, 16 Lac has been observed together with 
the $\beta$ Cephei variable 12 (DD) Lacertae. Results of the observations of 12 Lac and their analysis were published some years ago 
\citep*{HJR06}. In the next section, we describe the 2003-2004 multisite campaign's photometric observations and reductions. In 
Section 3, we carry out a frequency analysis of the campaign's $uvy$ time-series of 16 Lac. In Section 4 we use the analysis results 
to remove the intrinsic component of the variation from the time-series, thus bringing out the eclipse, and discuss the eclipse 
solutions and the evolutionary state of the secondary component. In Section 5, we derive the primary component's fundamental 
parameters. Section 6 is devoted to determining the harmonic degree of the three highest-amplitude pulsation modes of 16 Lac. The 
long-term variation of the photometric amplitudes and phases of these three modes is investigated in Section 7 using the present and 
archival data. The last section contains a summary and discussion of the results. The light variation of 2 And is examined in the 
Appendix. 

\section{Observations and reductions}

Our 2003-4 photometric observations of 16 Lac were carried out at ten observatories on three continents with small- to medium-sized 
telescopes (see Table 1). In most cases, single-channel differential photoelectric photometry was acquired. At the Sobaeksan Optical 
Astronomy Observatory (SOAO) and the Bia{\l}k\'ow Observatory (BO) the photometry was done with  CCD cameras. Wherever possible, 
Str\"omgren $uvy$ filters were used. At the Sierra Nevada and San Pedro Martir (SPM) Observatories simultaneous $uvby$ photometers 
were available, including $b$ filters as well. However, the $u$ data from SPM were unusable. At BO, a Str\"omgren $y$ filter was 
used. At four observatories where no Str\"omgren filters were available we used Johnson's $V$. Finally, as the photometer at the 
Mercator telescope had Geneva filters installed permanently, we used this filter system. We chose the two `classical' comparison 
stars: 10 Lac (O9V, $V ={}$4.88) and 2 And (A3Vn, $V ={}$5.09). A check star, HR\,8708 (A3Vm+F6V, $V ={}$5.81), was additionally 
observed during one of the SPM runs. HD\,216854 (F5, $V ={}$7.31) was used as a comparison star at SOAO; its constancy was checked 
against two fainter stars. At BO, BD\,+40\degr4950a = PPM\,63607 = GSC\,3223$-$01835 (F5/K0, $V ={}$9.29) was used as a sole 
comparison star. In order to compensate for the large brightness difference between the program and the comparison star ($\sim$3.7 
mag), BO observers took a sequence of CCD frames with short and long exposure-times, such that 16 Lac was well exposed on the frames 
with short exposure-times, while the comparison star, on the frames with long exposure-times. Reductions of the SOAO and BO 
observations included calibrating all CCD frames in a standard way and processing them with the Daophot II package \citep{s87}. 
Then, magnitudes were  obtained by means of aperture photometry, the magnitudes of the comparison stars were interpolated for the 
times of observation of 16 Lac and differential magnitudes were calculated. In the case of the BO data, several consecutive data 
points were averaged, resulting in the final time-series. 

\begin{table*}
  \caption{Log of the photometric measurements of 16 Lac. Observatories 
are listed in the order of their geographic longitude.}
  \begin{tabular}{@{}lrclrrcl@{}}
  \hline
  Observatory & Longitude & Latitude & Telescope & \multicolumn{2}{c}{Amount of data} 
& Filter(s) & Observer(s)\\
   &  &  &  & Nights & h\ \ \ \ &  \\
  \hline
 Sierra Nevada & $-$3\degr 23\arcmin & 37\degr 04\arcmin & 0.9-m & 18\ \ \ 
&  75.9 & $uvby$ & ER,PJA,RG  \\
Mercator & $-$17\degr 53\arcmin & 28\degr 46\arcmin & 1.2-m & 33\ \ \
& 104.2 & Geneva & KU,RD,JDD,TV,  \\
& & & & & & & JDR,BA,POB  \\
 Fairborn & $-$110\degr 42\arcmin & 31\degr 23\arcmin & 0.75-m APT & 
 57\ \ \  & 205.5 & $uvy$ & --- \\
 Lowell & $-$111\degr 40\arcmin & 35\degr 12\arcmin & 0.5-m & 19\ \ \  
& 96.8 & $uvy$ & MJ \\
 San Pedro Martir & $-$115\degr 28\arcmin & 31\degr 03\arcmin & 1.5-m & 10\ \ \ 
& 36.4 & $uvby$ & EP,JPS,LP \\
 Sobaeksan Optical & +128\degr 27\arcmin & 36\degr 56\arcmin & 0.6-m & 5\ \ \ 
& 31.9 & $V$ & SLK,JAL,SHK \\
 Astronomy &  &  &  &  &  &  & \\
 Mt.\ Dushak-Erekdag & +57\degr 53\arcmin & 37\degr 55\arcmin & 0.8-m & 12\ \ \ 
&63.4 & $V$ & TND,NID \\
 Mayaki & +30\degr 17\arcmin & 46\degr 24\arcmin & 0.5-m & 6\ \ \ 
&13.0 & $V$ & AIM \\
 Piszk\'estet\H o  & +19\degr 54\arcmin & 47\degr 55\arcmin & 0.5-m & 13\ \ \ 
& 56.1 & $V$ & MP,DZ,DL,VA \\
 Bia{\l}k\'ow & +16\degr 40\arcmin & 51\degr 29\arcmin & 0.6-m & 12\ \ \ 
& 67.2 &  $y$ & ZK,GM,JM\.Z,AP,MS \\
  \hline
Total & & & & 185\ \ \  & 749.3 & & \\
\hline
\end{tabular}
\end{table*}

As we mentioned in the Introduction, 2 And turned out to be a low-amplitude periodic variable.  We shall discuss the variability of 
2 And in the Appendix. No evidence for photometric variability of 10 Lac was found. We thus proceeded by pre-whitening the 
variability of 2 And with a fit determined from all its differential magnitudes relative to 10 Lac. The residual magnitudes of 2 And 
were then combined with the 10 Lac data into a curve that was assumed to reflect the effects of transparency and detector 
sensitivity changes only. These combined time-series were binned into intervals that would allow good compensation for the 
above-mentioned non-intrinsic variations in the target star time-series and were subtracted from the measurements of 12 and 16 Lac. 
Note that the binning minimizes the noise in the differential light curves of the targets. Finally, the photometric zero-points of 
the different instruments were compared and adjusted if required. In particular, adjustments were necessary for the SOAO and BO 
observations because they were obtained with different comparison stars. Measurements in the Str\"omgren $y$ and the Johnson and 
Geneva $V$ filters were analysed together because these filters have very nearly the same effective wave-length; the combined $y$, 
$V$ light-curve is henceforth referred to as `the $y$ data.' For further details of the reductions, common to 12 and 16 Lac, see 
\citet{HJR06}. 

\begin{figure*}
\includegraphics[angle=270,width=170mm]{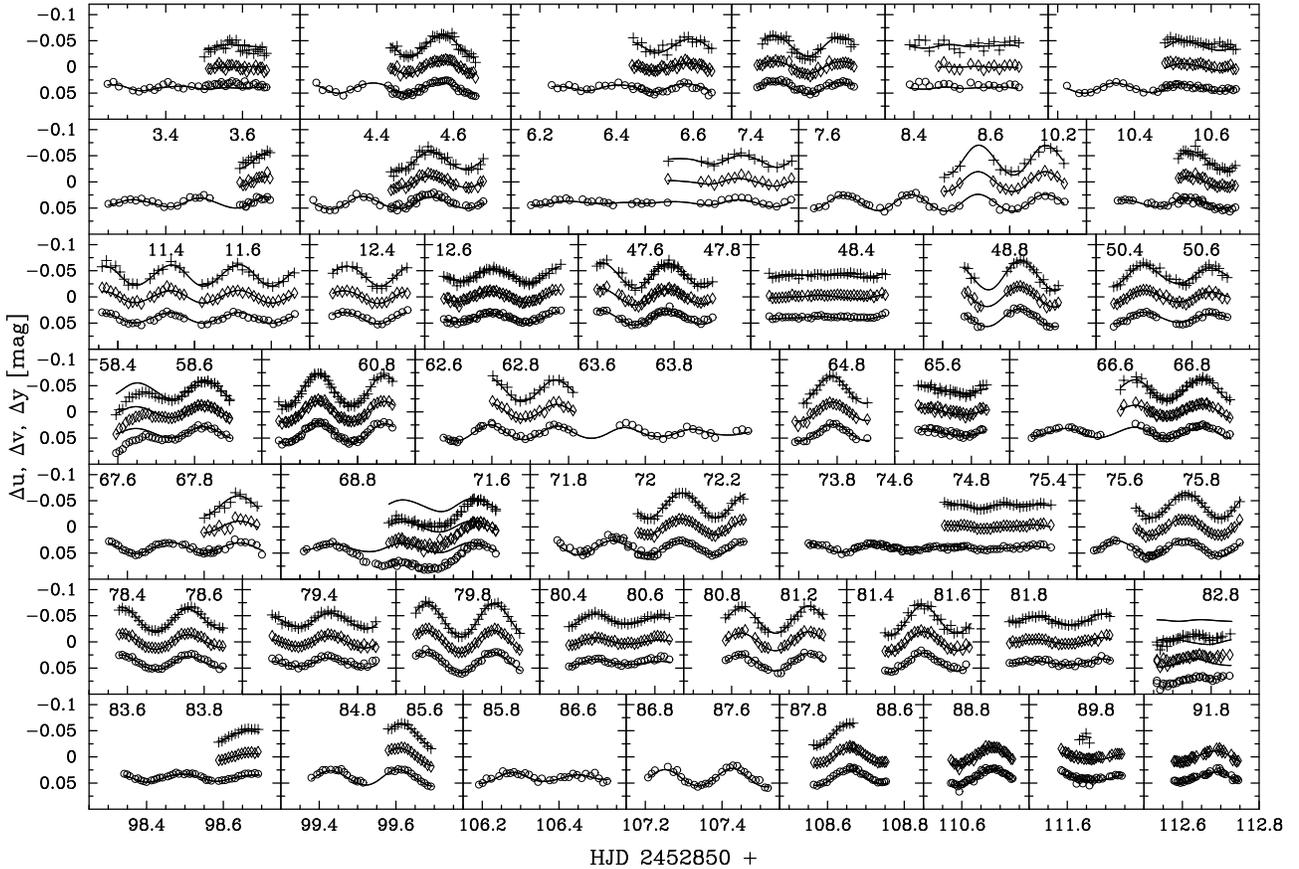}
\caption{A sample of our $u$, $v$, and $y$ light-curves of 16 Lac (plus signs, diamonds, and circles, respectively). The lines are 
the synthetic light-curves computed as explained in Section 3. Note the deviation of the measurements from the fit due to three 
consecutive eclipses on the nights JD 245 2917, JD 245 2929, and JD 245 2941. The amount of data displayed is about half the total.}
\label{Fig01}
\end{figure*}

For both stars, 12 and 16 Lac, the $y$ data were the most extensive by far. In the case of 16 Lac, there were 3190 $y$-filter  
measurements (including Johnson and Geneva $V$-filter measurements, see above), 2012 $v$-filter measurements, and 1686 $u$-filter 
measurements. The Mercator, Fairborn, Lowell, and Mayaki data included measurements falling between the first and the fourth 
contacts of six eclipses. Omitting these measurements resulted in the time-series suitable for frequency analysis; in the process, 
the number of measurements was reduced to 3055, 1895, and 1583 in $y$, $v$, and $u$, respectively. In all cases, the measurements 
spanned an interval of 179.2 d. The number of measurements in the Geneva filters was 411, with 395 taken outside eclipses. In both 
cases the data spanned 129.9 d. A sample of the light curves of 16 Lac is presented in Fig.~\ref{Fig01}.

\section{Frequency analysis}

Using the results of JP99's analysis (see the Introduction) we can predict blue-light amplitudes of the $f_i$ $(i ={}$1, 2, 3) modes 
for the epochs of our 2003-4 observations. In the case of $f_1$ and $f_2$, the amplitudes predicted under an assumption of 
sinusoidal variations with the reciprocal time-scales of 0.014 and 0.020 yr$^{-1}$ amount to 12.4 and 15.1 mmag, respectively, and 
both amplitudes should be very nearly constant over the 179.2-d interval spanned by the data. In the case of $f_3$, the amplitude 
should increase from 3.9 to 9.6~mmag over this interval. A preliminary analysis of the $y$ data showed that these predictions are in 
error, grossly so in the case of the first two modes. In order to examine this issue in detail, we divided the $y$, $v$, and $u$ 
data into adjacent segments; except for the first segments in all filters and the last two segments in $v$ and $u$, the segments 
partly overlapped. The first $y$ segment spanned 40 d and each of the remaining $y$, $v$, and $u$ segments, about 29 d. In each 
segment we then derived the amplitudes, $A_i$ $(i ={}$1, 2, 3), by the method of least squares using the following observational 
equations: 
\begin{equation} 
\Delta m_j = A_0 + \sum_{i=1}^N A_i \cos (2 \pi f_i t_j + \Phi_i), 
\end{equation} 
where $N ={}$3, $\Delta m_j$ are the $y$, $v$, or $u$ differential magnitudes, $f_i$ $(i ={}$ 1, 2, 3) are assumed to be equal to 
5.9112, 5.8550 and 5.5032 d$^{-1}$, respectively, and $t_j$ are HJD `minus' an arbitrary initial epoch. The assumed values of the 
frequencies $f_i$ $(i ={}$ 1, 2, 3) are approximately equal to the weighted means of the values given in Tables 4, 5 and 8, 
respectively, with the weights equal to the corresponding amplitudes (see Section 7). Note that the first $y$ segment covers about 2 
cycles of the longest beat-period in the variation of 16 Lac, the one arising from the interference of $f_1$ and $f_2$, and the 
remaining segments, about 1.5 cycle. An examination of the residuals from the least-squares solutions revealed six deviant 
observations in the second $y$ segment, all from the Mayaki Observatory. After rejecting these observations, the residuals in all 
but one segment fell within an interval of $\pm$0.015 mag; the exception was the first $u$ segment in which there were five 
residuals outside the $\pm$0.015 mag interval, but all smaller in their absolute value than 0.018 mag. 

\begin{figure} 
\includegraphics[width=76mm]{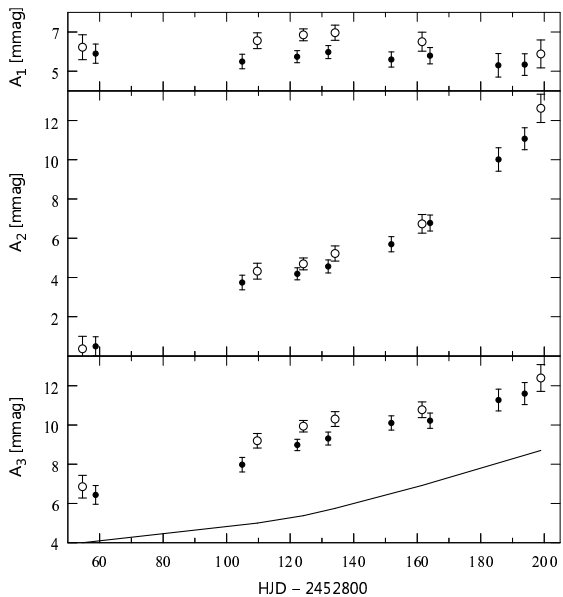} 
\caption{The $y$ (points) and $v$ (circles) amplitudes of the $f_1$, $f_2$, and $f_3$ modes of 16 Lac (from top to bottom) plotted 
as a function of HJD. The line in the bottom panel shows predicted blue-light amplitude of the $f_3$ mode (see the text for 
details).}
\label{Fig02}
\end{figure}

\begin{figure} 
\includegraphics[width=76mm]{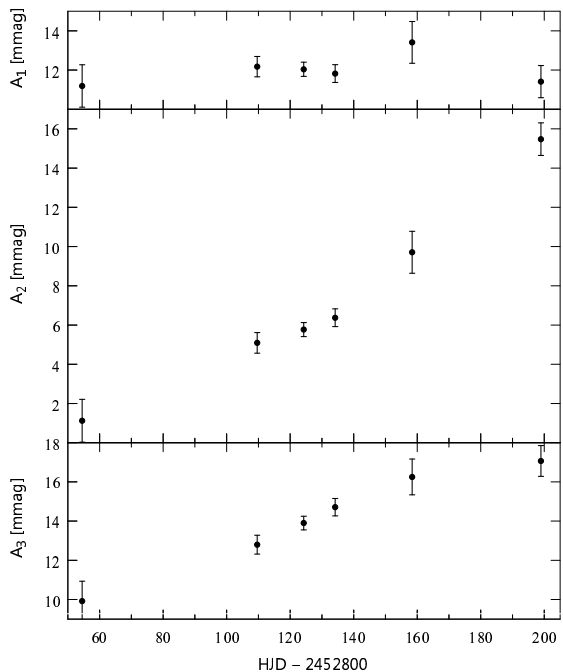} 
\caption{The $u$ amplitudes of the $f_1$, $f_2$, and $f_3$ modes of 16 Lac (from top to bottom) plotted as a function of HJD.}
\label{Fig03}
\end{figure}

Results of the above exercise are presented in Fig.~\ref{Fig02} for the $y$ and $v$ amplitudes (points and circles, respectively), 
and in Fig.~\ref{Fig03} for $u$. The error bars shown in the figures extend one standard deviation on each side of the plotted 
symbols. However, the standard deviations are not the formal standard deviations of the least-squares solutions of equations (1) but 
twice the formal standard deviations. In this, we follow \citet{H+00} and \citet{jhs} who---while dealing with time-series 
observations similar to the present ones---showed that the formal standard deviations were underestimated by a factor of about two. 

The solid line in the bottom panel of Fig.~\ref{Fig02} shows the predicted blue-light amplitude of the $f_3$ mode (see the first 
paragraph of this section). The predicted blue-light amplitude of the $f_1$ mode, 12.4 mmag, is a factor of about 2 greater than the 
observed $y$ and $v$ amplitudes seen in the top panel of Fig.~\ref{Fig02}. In the case of the $f_2$ mode, the prediction fails on 
two accounts: (1) the observed amplitudes vary from nearly zero to about 0.8 of the predicted value of 15.1 mmag on a time scale 
about 50 times shorter than that determined by JP99, and (2) the median values of the observed $y$ and $v$ amplitudes are a factor 
of about 2 smaller than predicted blue-light amplitude. 

\begin{figure} 
\includegraphics[width=76mm]{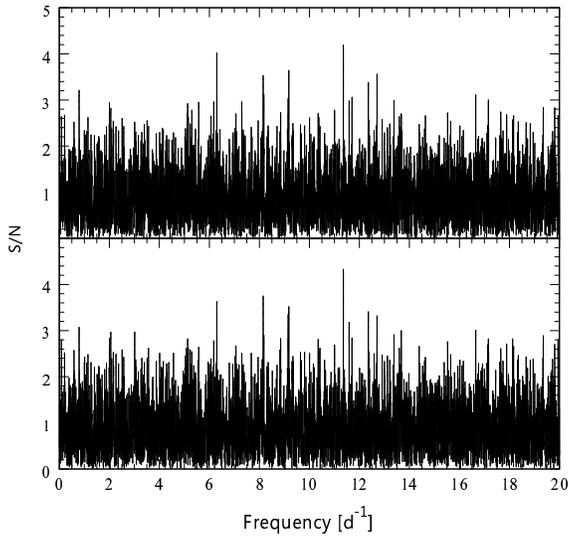} 
\caption{The signal-to-noise-ratio spectra, computed from the $y$ data pre-whitened with the frequencies $f_1 ={}$5.9112 d$^{-1}$, 
$f_2 ={}$5.8550 d$^{-1}$, and $f_3 ={}$5.5032 d$^{-1}$ by means of method 1 (upper panel) and method 2 (lower panel). For details, 
see text.}
\label{Fig04}
\end{figure}

Standard frequency analysis and successive pre-whitening with sinusoids applied to the 2003-4 $y$ magnitudes yielded the following 
frequencies (in the order they were identified from the power spectra): 5.503, 5.591, 5.856 and 5.852 d$^{-1}$. The first three 
numbers are very nearly equal to $f_3$, $f_1$ and $f_2$, while the fourth is close to $f_2$. The amplitudes amounted to 8.8, 5.9, 
4.3 and 1.8 mmag, respectively. The first two amplitudes are close to the mean $y$ amplitudes $A_3$ and $A_1$, while the sum of the 
third and the fourth is close to the mean $A_2$ (see Fig.~\ref{Fig02}). In addition, the fourth frequency differs from the third by 
less than the frequency resolution of the data. Clearly, the fourth frequency is an artefact. This shows that because of the 
variable amplitudes of the $f_2$ and $f_3$ modes, the usual procedure of pre-whitening with sinusoids is inadequate in the present 
case. Therefore, instead of the usual procedure, we tried the following two methods: (1) pre-whitening separately in each segment, 
(2) pre-whitening with $A_1$ assumed constant, $A_2$ assumed to vary quadratically with time, and $A_3$, linearly. Both methods were 
applied to the $y$ data; in the case of $v$ and $u$, we limited ourselves to method 1. In method 1, we computed residuals from the 
least-squares fit of equation (1) in each segment. Then, we took straight means of the residuals for a given $t_j$ in the 
overlapping parts of adjacent segments, so that for each $t_j$ there was one residual. In method 2, we used equation (1) with $A_1 = 
const$, $A_2 = B_1 + C_1 t_j + D_1 t_j^2$, and $A_3 = B_2 + C_2 t_j$. Since the first segment's $A_2$ would not fit the quadratic 
relation derived from the remaining segments (see the middle panel of Fig.~\ref{Fig02}), we applied the quadratic equation 
separately to this segment and then to the remaining data. In this case there was no need to average residuals because the first and 
second segment do not overlap. Using the residuals as data, we then computed the amplitude spectra and the signal to noise ratio 
($S/N$) as a function of frequency, where $S$ is the amplitude for a given frequency, and $N$ is the mean amplitude in 1 d$^{-1}$ 
frequency intervals; in the first frequency interval, we omitted the amplitudes for $f \le 1/T$, where $T$ is the total span of the 
data. For the $y$ residuals, the signal-to-noise ratios are plotted in Fig.~\ref{Fig04} as a function of frequency. We shall refer 
to the plots of this sort as the signal-to-noise spectra or $S/N$ spectra. In Fig.~\ref{Fig04}, the $S/N$ spectra are shown for 
method 1 and 2 in the upper and lower panel, respectively. In both cases, the highest $S/N$ peak occurs at 11.359 d$^{-1}$, a 
frequency very nearly equal to the combination frequency $f_2 + f_3$. The corresponding amplitudes amount to 0.56 and 0.59 mmag, and 
the $S/N$ values are equal to 4.3 and 4.5, respectively. Thus, both peaks are significant according to the popular criterion of 
\citet{B+93}. In the $S/N$ spectrum of method 1 residuals pre-whitened with $f_2 + f_3$, the highest peak, having the amplitude of 
0.64 mmag and $S/N ={}$4.2, occurred at 6.299 d$^{-1}$. In the analogous spectrum of method 2 residuals, there were no peaks with 
$S/N >{}$4; the peak at 6.299 d$^{-1}$ had $S/N ={}$3.4. The frequency of 6.299 d$^{-1}$ we shall refer to as $f_4$. In the $S/N$ 
spectrum of method 1 residuals pre-whitened with the five frequencies, there was a peak at 0.085~d$^{-1}$, rather close to $f_{\rm 
orb}$, but there was no peak at 2$f_{\rm orb}$, although one was present at this frequency in the power spectra of the 1965 $V$ data 
\citep[see figure 1 of][]{J93}. The amplitude and $S/N$ at $f_{\rm orb}$ were equal to 0.55 mmag and 2.1, respectively. At 2$f_{\rm 
orb}$, the corresponding numbers were 0.34 mmag and 1.3, respectively; the 1965 V amplitude was equal to 2.1 mmag.  

In the case of the $v$-filter residuals, computed using method 1, the $S/N$ spectrum showed the highest peak at 6.301 d$^{-1}$, very 
nearly equal to $f_4$. In the second $S/N$ spectrum, obtained from the residuals computed with this frequency included in 
pre-whitening, the highest peak occurred at 11.359 d$^{-1}$, the same combination frequency as that found in the $y$ residuals. The 
$S/N$ amounted to 3.8 and 3.9 in the first and the second spectrum, so that in the second spectrum the \citet{B+93} criterion of 
$S/N >{}$3.5 for a combination frequency was satisfied. The amplitudes were now greater than the $y$-filter ones, viz.\ 0.61 and 
0.64 mmag, respectively. At low frequencies, there were peaks close to $f_{\rm orb}$ and 2$f_{\rm orb}$. The $S/N$ and the amplitude 
at $f_{\rm orb}$ amounted to 2.7 and 0.66~mmag, while at 2$f_{\rm orb}$, to 2.4 and 0.58 mmag. The phase of maximum light of the 
$f_{\rm orb}$ term was 0.49$\,\pm\,$0.04 orbital phase, suggesting a reflection effect. In the case of the 2$f_{\rm orb}$ term, the 
phase of maximum light was 0.51$\,\pm\,$0.04 orbital phase, excluding an ellipsoidal variation as the cause.  The 1965 B amplitude 
at 2$f_{\rm orb}$ was equal to 1.6 mmag, and the phase of maximum light was 0.68$\,\pm\,$0.02 orbital phase \citep[see][table 
7]{J93}. In the case of the $u$-filter residuals, the highest peak in the first $S/N$ spectrum was at 12.359 d$^{-1}$, the 
$+$1~d$^{-1}$ alias of the combination frequency $f_2 + f_3$. In the second $S/N$ spectrum, the highest peak occurred at 
1.092~d$^{-1}$, and the second highest peak, at 6.301~d$^{-1}$. At 11.359 d$^{-1}$ in the first spectrum and at 6.301~d$^{-1}$ in 
the second spectrum, $S/N$ were equal to 3.9 and 3.1, respectively, and the $f_2 + f_3$ and $f_4$ amplitudes were equal to 0.82 and 
0.80 mmag. At $f_{\rm orb}$, the $S/N$ and the amplitude amounted to 2.0 and 0.75~mmag, while at 2$f_{\rm orb}$, to 1.0 and 0.38 
mmag. 

The $S/N$ spectrum of the $y$ residuals, computed by means of method 1 but with all five significant terms (i.e.\ $f_1$, $f_2$, 
$f_3$, $f_2+f_3$, and $f_4$) included showed no peaks higher than 3.8. We decided to terminate the frequency analysis at this stage. 
The $y$, $v$, and $u$ fits computed with the five terms taken into account were used to plot the synthetic light-curves in 
Fig.~\ref{Fig01}. 

\section{The eclipse}

\subsection{The {\sc EBOP} solutions}

\begin{figure} 
\includegraphics[width=76mm]{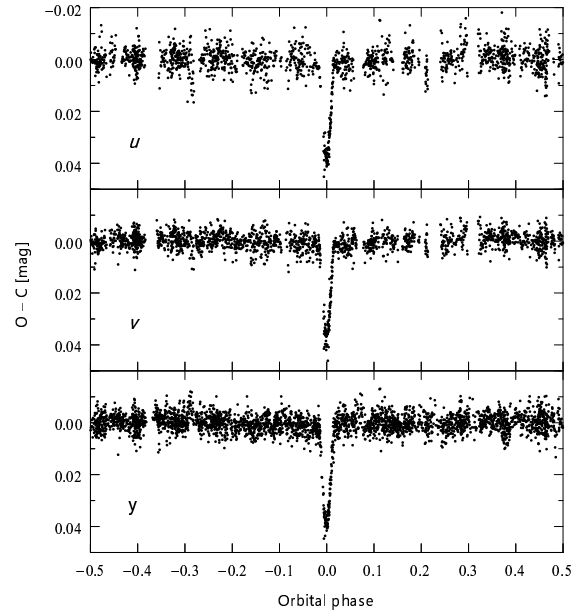} 
\caption{The $y$, $v$ and $u$ residuals from the five-term pulsational solutions plotted as a function of 
orbital phase. The $y$ deviant points (see Fig.~\ref{Fig06}) are not shown.}
\label{Fig05}
\end{figure}

\begin{figure} 
\includegraphics[width=76mm]{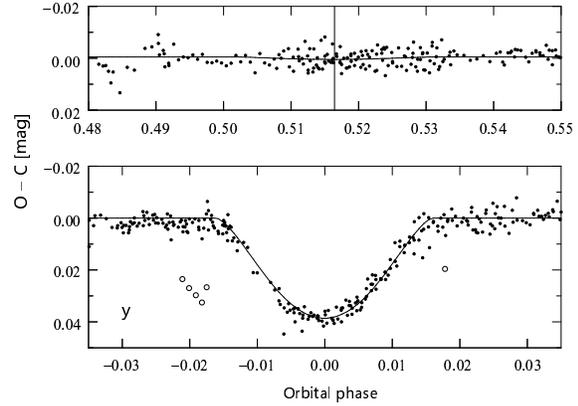} 
\caption{The $y$ residuals from the five-term pulsational solution in a limited interval of orbital phase around the primary and 
secondary eclipse. The vertical line in the upper panel indicates the predicted epoch of the secondary mid-eclipse. The other lines 
are fragments of the synthetic light-curve, computed as described in the text. The open circles are deviant points not included in 
the analysis.}
\label{Fig06}
\end{figure}

The residuals, computed by means of method 1 with the five terms taken into account (see the last paragraph of the preceding 
section) for all $y$, $v$, and $u$ observations, including those obtained during eclipses, are plotted in Fig.~\ref{Fig05} as a 
function of orbital phase. The ephemeris used in computing the phases was that of \citet{PJ88}, i.e. 
\begin{equation}
{\rm Min.\ light} = {\rm HJD}\,243\,9054.568 + 12.09684 E.
\end{equation}
In Fig.~\ref{Fig06} the $y$ residuals are shown in a limited range of orbital phase around the primary eclipse (lower panel) 
and those around the phase of the secondary mid-eclipse, predicted by the spectroscopic elements from solution IV of L01 (upper 
panel). No secondary eclipse can be detected: the mean $y$ residual in the $\pm$0.008 phase interval around the predicted 
mid-eclipse epoch amounts to 0.2$\,\pm\,$0.4 mmag. Clearly, the secondary component is much fainter than the primary. 

In an attempt to derive the parameters of 16 Lac we used \citet*{etz} computer program {\sc EBOP}. The program,  based on the 
Nelson-Davis-Etzel model \citep{nd,pe}, is well suited for dealing with detached systems such as the present one. The spectroscopic 
parameters $\omega$ and $e$, needed to run the program, were taken from solution IV of L01. The components were assumed to be 
spherical because the secondary component's mass is much smaller than that of the primary and the system is well detached: for any 
reasonable assumption about the primary component's mass, the mass ratio would be equal to about 0.13 and the semimajor axis of the 
very nearly circular relative orbit, to about 50 R$_{\sun}$ or 8 primary component's radii. However, we included reflected light 
from the secondary because a trace of a reflection effect can be seen in Fig.~\ref{Fig05}, especially in $y$ and $v$, in agreement 
with the results of the frequency analysis (see the fourth and the penultimate paragraphs of Section 3). The primary's 
limb-darkening coefficients were interpolated from table 2 of Walter Van 
Hamme\footnote{http://www2.fiu.edu/$\sim$vanhamme/limdark.htm, see also \citet{VH}}  for $T_{{\rm eff}} ={}$22\,500 K, $\log g 
={}$3.85 (see Section 5.1) and [M/H] = 0 \citep{Thou+, NDD05}. Unfortunately, since nothing is known about the secondary component 
except that it is much fainter than the primary, the central surface brightness of the secondary, $J_s$, which {\sc EBOP} uses as a 
fundamental parameter, must be derived indirectly. In units of the central surface brightness of the primary we have
\begin{equation}
J_{\rm s} = k^{-2}\, \frac{(1-u_{\rm p}/3)\, l_{\rm s}}{(1-u_{\rm s}/3)\, l_{\rm p}},
\end{equation}
where $l_p$ and $l_s$ are the normalized lights of the primary and the secondary, respectively, $u_p$ and $u_s$ are the limb 
darkening coefficients, and $k = r_s/r_p$ is the ratio of the radii. For $y$ we have
\begin{equation}
\log \frac{l_{\rm s}}{l_{\rm p}} = 2 \log k + 4 \log k_{\rm T} + \Delta BC/2.5,
\end{equation}
where $k_{\rm T} = T_{\rm eff,s}/T_{\rm eff,p}$ is the ratio of the effective temperatures of the components and $\Delta BC = 
BC_{\rm s} - BC_{\rm p}$ is the difference of the bolometric corrections. Introducing $l_{\rm s}/l_{\rm p}$ from this equation into 
equation (3) we get
\begin{equation}
J_{\rm s} =  \frac{1-u_{\rm p}/3}{1-u_{\rm s}/3}\,k^4_{\rm T} 10^{\Delta BC/2.5}\\[5mm]
\end{equation}
The bolometric correction of the primary component was taken from table 3 of \citet{Fl96} for $T_{\rm eff,p} ={}$22\,500 K. Assuming 
$T_{\rm eff,s}$, we read the secondary's bolometric correction from the same table as above.  Assuming further $\log g ={}$4.0 and 
[M/H] = 0, we read the secondary's limb-darkening coefficients from table 2 of Walter Van Hamme. Setting the integration ring size 
to 1\degr and the remaining parameters to their default {\sc EBOP} values, we run the program for several values of $T_{\rm eff,s}$ 
with $r_{\rm p}$, the relative radius of the primary component, $i$, the inclination of the orbit, $k$, the ratio of the radii, and 
$S_{\rm s}$, the reflected light from the secondary as unknowns. As data, we used the $y$ residuals (see Fig.~\ref{Fig05}, bottom 
panel). Unfortunately, we failed to find a solution which would converge. Convergent solutions were obtained if one of the first 
three unknowns, $r_{\rm p}$, $i$ or $k$, was fixed. After a number of trials we decided to fix $k$, leaving $r_{\rm p}$, $i$, and 
$S_{\rm s}$ as the unknowns. For a given $k$, identical triples of $r_{\rm p}$, $i$, and $S_{\rm s}$ were obtained for different 
$T_{\rm eff,s}$, i.e.\ different $J_{\rm s}$. For example, for $k ={}$0.23, $r_{\rm p}$ was equal to 0.1278$\,\pm\,$0.0015 and $i$ 
was 82\fdg93$\,\pm\,$0\fdg10, the same for $T_{\rm eff,s} ={}$5000 and 6050 K; $S_{\rm s}$ was equal to 0.00023$\,\pm\,$0.00005 for 
5000 K, and to 0.00025$\,\pm\,$0.00005 for 6050 K. The synthetic light-curves computed from these solutions were very nearly 
identical everywhere but around the secondary eclipse: the computed depth of the secondary eclipse was equal to 0.1 mmag for $T_{\rm 
eff,s} ={}$5000 K, while it was 0.7 mmag for $T_{\rm eff,s} ={}$6050~K. The mean residual in the $\pm$0.008 phase interval around 
the predicted mid-eclipse epoch amounted to 0.3$\,\pm\,$0.4 and $-$0.2$\,\pm\,$0.4 mmag for 5000 and 6050 K, respectively. The mean 
residual was equal to 0.0$\,\pm\,$0.4 mmag for 5750 K, and the computed depth of the secondary eclipse was then 0.5~mmag. For this  
computed depth of the secondary eclipse, the mean residual was equal to 0.0$\,\pm\,$0.4 mmag regardless of $k$. 

\subsection{Evolutionary state of the secondary component}

\begin{figure} 
\includegraphics[width=76mm]{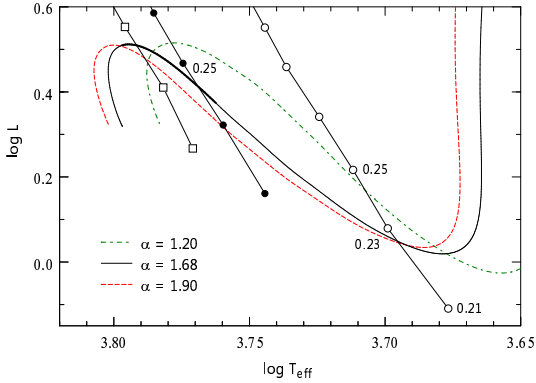} 
\caption{The HR-diagram positions of the secondary component of 16 Lac for three series of eclipse solutions, obtained with the 
ratio of the radii $k = r_{\rm s}/r_{\rm p}$ as the parameter and assuming the depth of the secondary eclipse to be equal to 0.1 
mmag (open circles), 0.5 mmag (dots) and 0.7 mmag (open squares). The open circles correspond to $k$ from 0.21 to 0.31 with a step 
of 0.02, the dots, to $k$ from 0.21 to 0.27 with a step of 0.02, and the open squares, to $k$ equal to 0.21, 0.23, and 0.25; three 
open circles and one dot are labeled with their values of $k$. Also shown are Pisa 1.3~M$_{\sun}$ pre-MS evolutionary tracks for 
three values of the mixing-length parameter $\alpha$ (the solid line, the green dash-dotted line and the red dashed line). The 
thickened segment of the $\alpha ={}$1.68 track indicates the age of 16.3$\,\pm\,$1.3 Myr, i.e.\ the evolutionary age of the primary 
component (see Section 5.2).} 
\label{Fig07} 
\end{figure}

For a range of $k$, synthetic light-curves computed from the solutions with an assumed depth of the secondary eclipse are 
indistinguishable from one another. Each solution yields an abscissa for plotting the secondary component in the HR diagram. For an 
assumed mass of the primary component, one can also have the secondary's radius in absolute units from $r_{\rm p}$, $i$ and the 
spectroscopic elements $K_1$ and $e$, and therefore, the secondary's ordinate in the HR diagram. For three values of the assumed 
depth of the secondary eclipse, 0.1, 0.5 and 0.7 mmag, and the primary component's mass $M_{\rm p} ={}$10 M$_{\sun}$, positions of 
the secondary component in the HR diagram are shown in Fig.~\ref{Fig07} for a range of $k$. 

For the range of $k$ shown in Fig.~\ref{Fig07}, the assumption of $M_{\rm p} ={}$10 M$_{\sun}$ implies 1.30${}\leq M_{\rm s} 
<{}$1.31 M$_{\sun}$. Therefore, without making noticeable errors, we can compare the HR-diagram positions of the secondary component 
with $M ={}$1.30 M$_{\sun}$ evolutionary tracks. The evolutionary tracks plotted in the figure are the 1.30 M$_{\sun}$, $Y 
={}$0.265, $Z ={}$0.0175 Pisa pre-main-sequence (pre-MS) tracks (\citealt*{Tog+11}, http://astro.df.unipi.it/stellar-models/). The 
tracks were computed using the mixing-length theory of convection with three values of the mixing length $l = \alpha H_{\rm p}$, 
where $\alpha$ is the mixing-length parameter and $H_{\rm p}$ is the pressure scale height. The mid-value of $l$ ($\alpha ={}$1.68) 
was calibrated by means of the Pisa standard solar model \citep[for further details, see][]{Tog+11}. The thickened segment of the 
$\alpha ={}$1.68 track indicates the age within 1\,$\sigma$ of 16.3 Myr, the evolutionary age of the primary component (see Section 
5.2). Since the duration of the pre-MS phase of the primary component's evolution is of the order of 0.1 Myr, we conclude that the 
secondary component is in the pre-MS contraction phase, the same conclusion as that reached long ago by \citet{PJ88}. Now, the 
points of intersection of the three series of the eclipse solutions in Fig.~\ref{Fig07} with the pre-MS tracks constrain the ratio 
of the radii to 0.23${}\la k \la{}$0.27. If the mass of the primary component were assumed to be equal to $M_{\rm p} ={}$8.8 
M$_{\sun}$, corresponding to $M_{\rm s} \approx{}$1.20 M$_{\sun}$, the constraints would be 0.21${}\la k \la{}$0.25. If $M_{\rm p} 
={}$11.2 M$_{\sun}$ ($M_{\rm s} \approx{}$1.40 M$_{\sun}$), 0.23${}\la k \la{}$0.29. Thus, for 8.8${}\leq M_{\rm p} \leq{}$11.2 
M$_{\sun}$ we get 0.21${}\la k \la{}$0.29. Over this range of $k$, the relative radius of the primary component, $r_{\rm p}$, is a 
monotonically increasing function of $k$, while the inclination of the orbit, $i$, is monotonically decreasing with $k$, and both 
are virtually independent of the assumed depth of the secondary eclipse. Thus, from the last inequality we have 0.125${}\la r_{\rm 
p} \la{}$0.132 and 83\fdg4${}\ga i \ga{}$82\fdg0. More importantly, we can also obtain the lower and upper bound of the logarithmic 
surface gravity of the primary component: 3.78${}\la \log g_{\rm p} \la{}$3.87. The formal standard deviations of $r_{\rm p}$, $i$ 
and $\log g_{\rm p}$, equal to 0.0013, 0\fdg10 and 0.011 dex, respectively, are---not surprisingly---much smaller than the allowed 
ranges of $r_{\rm p}$, $i$ and $\log g_{\rm p}$. 

The eclipse solutions which predict the depth of the secondary eclipse to be equal to 0.5 mmag yield synthetic light-curves which 
best fit the data around the secondary eclipse (see the end of the last paragraph of Section 4.1) while they fit the data elsewhere 
as well as do the other solutions. The fact that the corresponding line in Fig.~\ref{Fig07} (the solid line with dots) crosses the 
$\alpha ={}$1.68 evolutionary track at an evolutionary age within the range of the evolutionary age of the primary component (see 
Section 5.2) is encouraging. It would be worthwhile to carry out space photometry of 16 Lac in order to find out whether the depth 
of the secondary eclipse is indeed close to 0.5 mmag. In any case, space photometry will be necessary to better constrain the 
range of $k$, and therefore the ranges of $r_{\rm p}$, $i$ and $\log g_{\rm p}$.

\section{Fundamental Parameters}

\subsection{The effective temperature and the surface gravity}

The effective temperature and surface gravity of 16 Lac can be obtained from the Str\"omgren indices using several photometric 
calibrations available in the literature. The $c_1$ index from \citet{hm}, corrected for the interstellar reddening in the standard 
way \citep{c78}, yielded the following values of $T_{\rm eff}$ (with the calibration referenced in the parentheses after each 
value): 22\,435 K \citep{ds77}, 22\,580 K ({\sc UVBYBETA}\footnote{A FORTRAN program based on the grid published by \citet{md85}. 
Written in 1985 by T.T.\ Moon of the University London and modified in 1992 and 1997 by R.\ Napiwotzki of Universitaet Kiel 
\citep*[see ][]{n93}.}), 22\,430 K \citep{SJ93}, and 22\,635 K \citep{b94}. In the case of the \citet{b94} calibration, the $\beta$ 
index was also needed in addition to $c_0$. Taking a straight mean of these values we get $T_{\rm eff} ={}$22\,520 K, with a formal 
standard error equal to 50 K. The latter number is so small because the four photometric calibrations are not independent; they all 
rely heavily on the OAO-2 absolute flux calibration of \citet{cod}. Realistic standard deviations of the effective temperatures of 
early-type stars, estimated from the uncertainty of the absolute flux calibration, amount to about 3\,\% \citep{n93,j94} or 680 K 
for the $T_{\rm eff}$ in question. Thus, $T_{\rm eff}$ of 16 Lac, obtained from the Str\"omgren indices, is equal to 
22\,520$\,\pm\,$680 K. The most recent spectroscopic determinations of $T_{\rm eff}$ include 22\,900$\,\pm\,$1000 K \citep{Thou+}, 
21\,500$\,\pm\,$750 K \citep*{Prug+} and 23\,000$\,\pm\,$200 K \citep{NP12}. A straight mean of these numbers is equal to 
22\,470$\,\pm\,$480 K, in surprisingly good agreement with the photometric value. We shall adopt 22\,500$\,\pm\,$600 K as the 
$T_{\rm eff}$ of 16 Lac. 

The logarithmic surface gravity of 16 Lac derived from $\beta$ and $c_0$ turned out to be equal to 3.93 ({\sc UVBYBETA}) and 3.90 
\citep{b94}. The good agreement between these values may be misleading: according to \citet{n93}, the uncertainty of photometric 
surface gravities of hot stars is equal to 0.25 dex. We conclude that the photometric $\log g$ of 16 Lac is equal to 
3.90$\,\pm\,$0.25. The spectroscopic values of $\log g$ are equal to 3.80$\,\pm\,$0.20 \citep{Thou+}, 3.75$\,\pm\,$0.17 
\citep{Prug+} and 3.95$\,\pm\,$0.05 \citep{NP12}. A straight mean of these numbers is equal to 3.83, with a standard error equal to 
0.06 dex. We shall adopt 3.85$\,\pm\,$0.15 as the $\log g$ of 16 Lac, where the adopted standard deviation, equal to the median of 
the standard deviations of the individual $\log g$ values, is a compromise between the standard deviation of the photometric $\log 
g$ and that of the spectroscopic $\log g$ of \citet{NP12}. Note that because of negligible brightness of the secondary component, 
the $T_{\rm eff}$ and $\log g$ we adopted pertain to the primary component. 

\subsection{The effective temperature -- surface gravity diagram}

\begin{figure} 
\includegraphics[width=76mm]{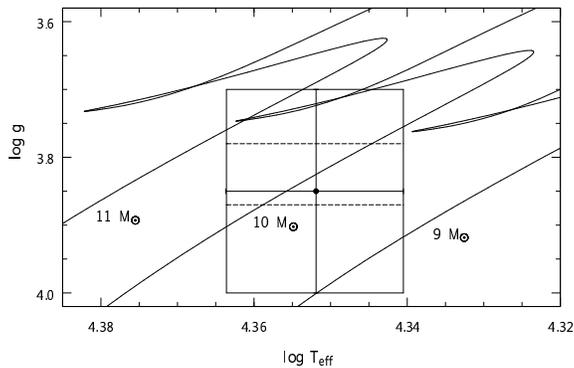} 
\caption{The primary component of 16 Lac plotted in the effective temperature -- surface gravity plane using $T_{\rm eff}$ and  
$\log g$ from Section 5.1 (dot with error bars).  The dashed horizontal lines indicate the lower and upper bound of $\log g_{\rm 
p}$, obtained in Section 4.2 from the eclipse solutions and pre-MS evolutionary tracks. The 9, 10 and 11 M$_{\sun}$ evolutionary 
tracks are explained in the text.} 
\label{Fig08} 
\end{figure}

In Fig.~\ref{Fig08}, the primary component of 16 Lac is plotted in the $\log T_{\rm eff}$ -- $\log g$ plane using the effective 
temperature and surface gravity from Section 5.1 (dot with error bars). The dashed horizontal lines indicate the lower and upper 
bound of $\log g_{\rm p}$, obtained in Section 4.2 from the eclipse solutions. Also shown are evolutionary tracks computed by means 
of the Warsaw - New Jersey evolutionary code \citep[see, e.g.,][]{P+98}, assuming the initial abundance of hydrogen X = 0.7 and the 
metallicity Z = 0.015, the OPAL equation of state \citep{RN02} and the OP opacities \citep{S05} for the latest heavy element mixture 
of \citet{A+09}. For lower temperatures, the opacity data were supplemented with the Ferguson tables \citep{FAA05, S+09}. We assumed 
no convective-core overshooting and $V_{\rm rot} ={}$ 20 km~s$^{-1}$ on the zero-age main sequence, a value consistent with the 
observed $V_{\rm rot} \sin i$ \citep*{GGS} under the assumption that the rotation and orbital axes are aligned. The effect of 
rotation on $\log g$ was taken into account by subtracting the centrifugal acceleration, amounting in the present case to about 
0.001~dex. 

The evolutionary mass at the position of the dot in Fig.~\ref{Fig08}, $M_{\rm ev} ={}$9.8$\,\pm\,$1.3 M$_{\sun}$; the standard 
deviation of $\log g$ is responsible for the standard deviation of $M_{\rm ev}$. If the lower and upper bound of $\log g_{\rm p}$ 
were used, the result would be 9.3${}\la M_{\rm ev} \la{}$10.8 M$_{\sun}$. According to \citet{NP12}, $M_{\rm ev} 
={}$9.8$\,\pm\,$0.3 M$_{\sun}$. In this case, the small standard deviation of $M_{\rm ev}$ is the consequence of the small standard 
deviations these authors assign to their spectroscopic $T_{\rm eff}$ and $\log g$. The evolutionary age of 16 Lac, obtained from 
Fig.~\ref{Fig08}, amounts to 16.3$\,\pm\,$1.5 Myr, in surprisingly good agreement with the \citet{B64} estimate of the age of Lac 
OB1a. The most recent asteroseismic analysis of the star \citep{Thou+} has led to a mass of 9.62$\,\pm\,$0.11 M$_{\sun}$ and an age 
of 15.7 Myr. Clearly, the accuracy of the asteroseismic values is much higher than that attainable by either the photometric or 
spectroscopic method. However, the sensitivity of the asteroseismic values to the details of modeling needs to be examined. We 
leave this for a future paper.  

\section{The harmonic degree of the three highest-amplitude modes}

\subsection{From the $uvy$ data}

Our data are the most extensive photometric observations of 16 Lac ever obtained. In addition, the three photometric passbands we 
used include one on the short-wavelength side of the Balmer jump and two in the Paschen continuum. Thus, we can derive the amplitude 
ratios, $A_y/A_u$ and $A_v/A_u$, and the phase differences, $\Phi_y-\Phi_u$ and $\Phi_v-\Phi_u$, which will be more accurate than 
any available before and sensitive to the harmonic degree of the pulsation modes. However, before computing the amplitudes and the 
phases we had to tackle the problem of the differences in the number and the time distribution between the $y$, $v$ and $u$ data. 
Since the $u$ data are fewer in number and less evenly distributed in time than the $y$ data, there were fewer $u$ segments than the 
$y$ segments (compare Fig.~\ref{Fig03} with Fig.~\ref{Fig02}). Consequently, the mean epochs of the $u$ segments did not match those 
of the $y$ segments. To a smaller degree, this was also the case with the $v$ data. Because the amplitudes vary from one segment to 
another, the amplitude ratios computed using amplitudes from unmatched segments would be biased; the same goes for the phases and 
the phase differences. We therefore divided the $y$ and $v$ data into new segments, in most cases different from those we formed in 
Section 3. In the new segments, the initial and final epochs and the time distribution of the data matched those of the $u$ segments 
as closely as the data allowed. Then, the amplitudes and the phases in each segment were derived by fitting equation (1) with $N 
={}$5 to the data; the frequencies $f_i$ $(i ={}$1,..,$4$) and $f_5 = f_2 + f_3$ were the same as in Section 3. Finally, the 
amplitudes of the three highest-amplitude modes from the matching segments were used to compute the amplitude ratios, and the 
phases, to compute the phase differences. In spite of the high quality of our photometry, standard deviations of the phase 
differences were rather large, rendering them useless for mode identification. The amplitude ratios and their weighted means are 
listed in Table 2. There is no evidence in the table for a time-variability of the amplitude ratios, a result consistent with the 
fact that we are dealing with normal pulsation modes. In computing the weighted means, we assumed weights inversely proportional to 
the squares of the standard deviations of the components. Standard deviations of the weighted means were computed by adding the 
standard deviations of the components in quadrature and dividing the sum by the number of components. Note that for $f_2$, the 
JD\,245\,2855.4 values are deviant. This is because at this epoch the $f_2$ amplitudes were close to zero. In computing the weighted 
means, we omitted the JD\,245\,2855.4 values. 

\begin{table*}
  \caption{The $uy$ and $uv$ amplitude ratios for the three highest-amplitude modes. In the case of $f_2$, the 
HJD\,245\,855.4 values were omitted in computing the weighted means.}
  \begin{tabular}{@{}ccccccc@{}}
  \hline
   & \multicolumn{2}{c}{$f_1$} & \multicolumn{2}{c}{$f_2$}& \multicolumn{2}{c}{$f_3$}\\
HJD-245\,2800&$A_y/A_u$&$A_v/A_u$&$A_y/A_u$&$A_v/A_u$&$A_y/A_u$&$A_v/A_u$ \\
  \hline
\hspace{3pt}55.4 &0.565$\pm$0.078&0.560$\pm$0.082&0.324$\pm$0.870&0.270$\pm$0.900&0.643$\pm$0.088&0.697$\pm$0.096\\
107.1 &0.494$\pm$0.040&0.540$\pm$0.040&0.697$\pm$0.108&0.843$\pm$0.114&0.649$\pm$0.040&0.723$\pm$0.040\\
124.8 &0.489$\pm$0.028&0.567$\pm$0.030&0.753$\pm$0.066&0.816$\pm$0.072&0.655$\pm$0.026&0.719$\pm$0.028\\
135.1 &0.505$\pm$0.034&0.588$\pm$0.040&0.743$\pm$0.072&0.831$\pm$0.084&0.647$\pm$0.028&0.706$\pm$0.034\\
160.0 &0.444$\pm$0.050&0.494$\pm$0.056&0.663$\pm$0.084&0.676$\pm$0.088&0.643$\pm$0.042&0.685$\pm$0.044\\
198.4 &0.440$\pm$0.060&0.517$\pm$0.070&0.758$\pm$0.056&0.818$\pm$0.060&0.676$\pm$0.046&0.723$\pm$0.052\\
  \hline
Wt.~Mean = &0.488$\pm$0.021&0.554$\pm$0.023&0.734$\pm$0.035&0.800$\pm$0.038&0.652$\pm$0.020&0.712$\pm$0.022\\ 
\hline
\end{tabular}
\end{table*}

\begin{figure}
\includegraphics[width=80mm]{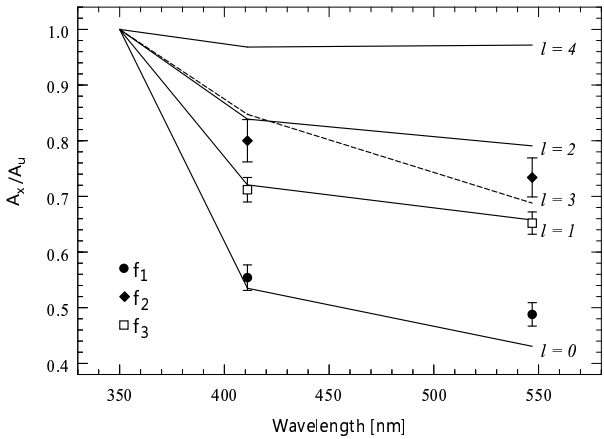} 
\caption{A comparison of the observed (symbols with error bars) and theoretical (lines) $uvy$ amplitude ratios for the three 
highest-amplitude pulsation modes of 16 Lac, $f_1$, $f_2$ and $f_3$. The observed amplitude ratios are the weighted means listed in 
the bottom line of Table 2. The theoretical amplitude ratios correspond to the dot in Fig.~\ref{Fig08}. The theoretical $\ell ={}$3 
amplitude ratios are shown with the dashed line.}
\label{Fig09}
\end{figure}

A comparison of the observed amplitude ratios with the theoretical ones is presented in Fig.~\ref{Fig09}. The theoretical amplitudes 
were computed according to the zero-rotation formulae of \citet{DD+02} using the nonadiabatic pulsational code of \citet{D77} and 
Kurucz line-blanketed LTE model atmospheres \citep{Ku04} for [M/H] = 0.0 and the microturbulent velocity $\xi = 2$~km/s. The 
remaining input parameters were the same as those used in computing the evolutionary tracks in Section 5.2. The calculations were 
carried out for $\log T_{\rm eff}$ and $\log g$ used to plot the dot in Fig.~\ref{Fig08}. The $\log g_{\rm p}$ obtained in Section 
4.2, although more precise than $\log g$, was not used because it may be less accurate on account of being model-dependent. As can 
be seen from Fig.~\ref{Fig09}, $f_1$ should be identified with a radial mode, notwithstanding that the agreement between the 
observed and theoretical $A_y/A_u$ is problematic. The remaining two modes are nonradial with $\ell \le{}$3 because neither the $\ell 
={}$0, nor the $\ell ={}$4  line fits their amplitude ratios. In the case of $f_3$, the observed and theoretical amplitude ratios 
agree to within 1\,$\sigma$ for $\ell ={}$1, while in the case of $f_2$, the observed $A_v/A_u$ falls about 1\,$\sigma$ below the 
$\ell ={}$2 and 3 lines while the observed $A_y/A_u$ lies half-way between the $\ell ={}$2 and 3 lines. 

\begin{figure} 
\includegraphics[width=80mm]{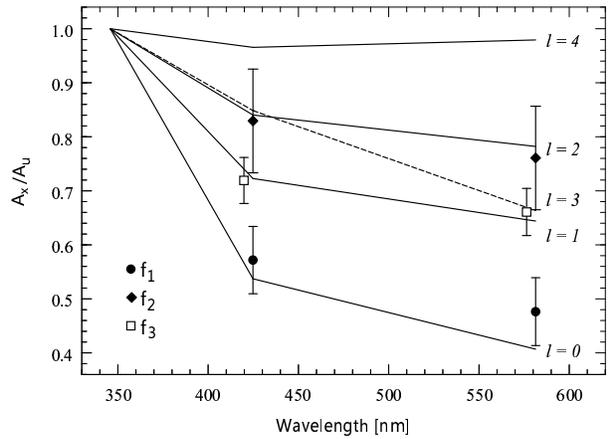} 
\caption{A comparison of the observed (symbols with error bars) and theoretical (lines) $UBG$ amplitude ratios for the three 
highest-amplitude pulsation modes of 16 Lac, $f_1$, $f_2$ and $f_3$; the open square was shifted slightly along the abscissa to 
avoid overlap. The observed amplitude ratios are the weighted means listed in the bottom line of Table 3. The theoretical amplitude 
ratios correspond to the dot in Fig.~\ref{Fig08}.}
\label{Fig10}
\end{figure}

\subsection{From the Geneva $UBG$ data}

\begin{table*}
  \caption{The Geneva $UG$ and $UB$ amplitude ratios for the three highest-amplitude modes.}
  \begin{tabular}{@{}ccccccc@{}}
  \hline
   & \multicolumn{2}{c}{$f_1$} & \multicolumn{2}{c}{$f_2$}& \multicolumn{2}{c}{$f_3$}\\
HJD-245\,2800&$A_G/A_U$&$A_B/A_U$&$A_G/A_U$&$A_B/A_U$&$A_G/A_U$&$A_B/A_U$ \\
  \hline
135.7 &0.477$\pm$0.086&0.580$\pm$0.086&0.727$\pm$0.155&0.821$\pm$0.158&0.647$\pm$0.060&0.721$\pm$0.060\\
181.9 &0.476$\pm$0.092&0.563$\pm$0.090&0.778$\pm$0.112&0.833$\pm$0.108&0.676$\pm$0.063&0.717$\pm$0.060\\
  \hline
Wt.~Mean = &0.476$\pm$0.063&0.572$\pm$0.062&0.761$\pm$0.096&0.829$\pm$0.096&0.661$\pm$0.044&0.719$\pm$0.043\\ 
\hline
\end{tabular}
\end{table*}

The data obtained with the Geneva filters (see Section 2) cover three intervals: JD\,245\,2861.6 to JD\,245\,2872.7, JD\,245\,2921.5 
to JD\,245\,2949.5, and JD 245\,2971.3 to JD\,245\,2991.5. From the data in the latter two intervals we derived the amplitude 
ratios, $A_G/A_U$ and $A_B/A_U$ in the same way as from the $uvy$ data in Section 6.1. The amplitude ratios are listed in Table 3. 
The weighted mean amplitude ratios are plotted as a function of the passbands' central frequency in Fig.~\ref{Fig10}. Also plotted 
are the theoretical amplitude ratios. The harmonic-degree identification of $\ell ={}$0 for $f_1$ and $\ell ={}$1 for $f_3$ inferred 
in Section 6.1 from the $uvy$ data is confirmed. In the case of $f_2$, $\ell ={}$2 fits now better than $\ell ={}$3. In addition, 
$\ell ={}$3 would be less satisfactory than $\ell ={}$2 because of the effect of cancellation in integrating over the stellar disc. 
However, the standard deviations of the $f_2$ amplitude ratios are rather large.

These harmonic-degree identifications, i.e.\ $\ell ={}$0 for $f_1$, $\ell ={}$2 or, less satisfactorily, $\ell ={}$3 for $f_2$, and 
$\ell ={}$1 for $f_3$, agree with the earlier identifications, based on the $UBV$ amplitude ratios and the $V$-amplitude to the 
RV-amplitude ratio \citep[see][]{DJ96}. They have points in common with the spectroscopic identifications of \citet{ALB+03a} and 
\citet{ALB+03b}. An analysis of the line profiles of the He\,I $\lambda$6678 {\AA} line led \citet{ALB+03a} to the conclusion that 
$f_1$ should be identified with a radial mode, $f_2$, with an $\ell ={}2$, $m ={}$0 mode, and $f_3$, with an $\ell ={}$1, $m ={}$0 
mode. Subsequently, \citet{ALB+03b} modified the harmonic-degree identification for $f_3$ to $\ell <{}$3. For $f_2$, the 
identification of \citet{ALB+03b} is thus more specific than ours, while the reverse is true for $f_3$. 

\section{Long-term variation of the amplitude and phase of the large-amplitude terms}

\subsection{The $f_1$ term}

All out-of-eclipse blue-filter observations of 16 Lac obtained throughout 1992 span an interval of over 40 years and consist of 6 
334 data points (see JP99). By supplementing these data with our out-of-eclipse $y$ observations we formed a data set of 9 384 
points, spanning an interval of 53.4 years. We shall refer to this set as $B\&y$. Subtracting the contribution of the $f_i$ and 
$f_j$ terms ($i,j ={}$1, 2, 3, $i \neq j$) from $B\&y$ resulted in three sets which we shall refer to as $B\&y -{}$23, $B\&y -{}$13 
and $B\&y -{}$12. We chose the archival blue-filter observations and the present $y$ observations because they are much more 
numerous than observations in other filters. In the following frequency analysis of the combined data we shall neglect the 
difference between the blue and yellow pulsation amplitudes. This will lead to some amplitude smearing in the results reported in 
this and the two following sections. In 1965, when the amplitudes of the three high-amplitude terms were close to their maximum 
values, the difference between the $B$ and $V$ amplitudes amounted to 2.0$\,\pm\,$0.21, 0.40$\,\pm\,$0.21 and 0.70$\,\pm\,$0.20 mmag 
for $f_1$, $f_2$ and $f_3$, respectively \citep[see][tables 7 and 8]{J93}. Thus, the amplitude smearing will be the largest (albeit 
far from severe) in the case of $f_1$ and very nearly negligible in the remaining cases. 

\begin{figure} 
\includegraphics[width=76mm]{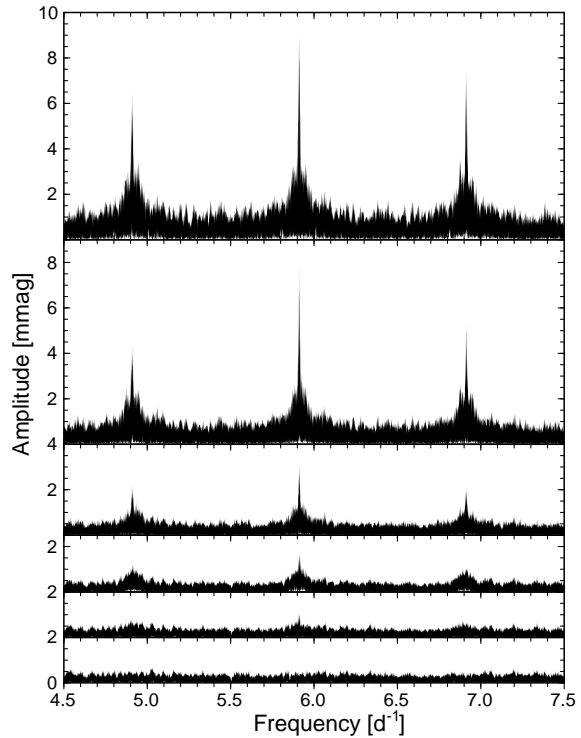} 
\caption{The amplitude spectra of $B\&y -{}$23, i.e.\ the 1950-1992 archival blue-filter data and our $y$-filter data combined, 
freed from the contribution of the $f_2$ and $f_3$ terms. Starting with the second panel from the top, the data were pre-whitened 
with (1) $f_{1,1} ={}$5.911312 d$^{-1}$, (2) $f_{1,1}$ and $f_{1,2}={}$5.911275 d$^{-1}$, (3) $f_{1,1}$, $f_{1,2}$ and 
$f_{1,3}={}$5.911350 d$^{-1}$, (4) $f_{1,1}$, $f_{1,2}$, $f_{1,3}$ and $f_{1,4}={}$5.911230~d$^{-1}$, and (5) $f_{1,1}$, $f_{1,2}$, 
$f_{1,3}$, $f_{1,4}$ and $f_{1,5}={}$5.911395 d$^{-1}$.}
\label{Fig11}
\end{figure}

The amplitude spectra of $B\&y -{}$23 are shown in Fig.~\ref{Fig11}. The abscissae of the highest peaks in the amplitude spectra 
are given in the caption to the figure and are listed in the second column of Table 4; the corresponding periods are given in column 
three. The amplitudes and phases with their formal standard deviations, obtained from a five-frequency least-squares fit of equation 
(1) to $B\&y -{}$23 are listed in columns four and five. The standard deviation of the fit amounted to 5.4 mmag. The epochs of 
observations were reckoned from HJD\,244\,5784. 

\begin{table}
 \centering
 \begin{minipage}{80mm}
\caption{Fine structure of the $f_1$ term.}
\begin{center}
  \begin{tabular}
{rccr@{\hspace{3pt}$\pm$\hspace{3pt}}lr@{\hspace{3pt}$\pm$\hspace{3pt}}l} \\
  \hline
  $j$ & \multicolumn{1}{c}{$f_{1,j}$ [d$^{-1}$]}& \multicolumn{1}{c}{$P_{1,j}$ [d]} & \multicolumn{2}{c}{$A_{1,j}$ [mmag]} & 
\multicolumn{2}{c}{$\Phi_{1,j}$ [rad]} \\
 \hline
  1&\hspace{1pt} 5.911312&0.1691672 &12.45 & 0.09& 0.181& 0.007\\
  2&\hspace{1pt} 5.911275&0.1691682 & 9.06 & 0.09& 4.080& 0.010\\
  3&\hspace{1pt} 5.911350&0.1691661 & 3.27 & 0.09& 3.326& 0.027\\
  4&\hspace{1pt} 5.911230&0.1691695 & 2.26 & 0.09& 1.836& 0.039\\
  5&\hspace{1pt} 5.911395&0.1691648 & 1.35 & 0.09& 5.367& 0.065\\
\hline
\end{tabular}
\end{center}
\end{minipage}
\end{table}

\begin{figure} 
\includegraphics[width=85mm]{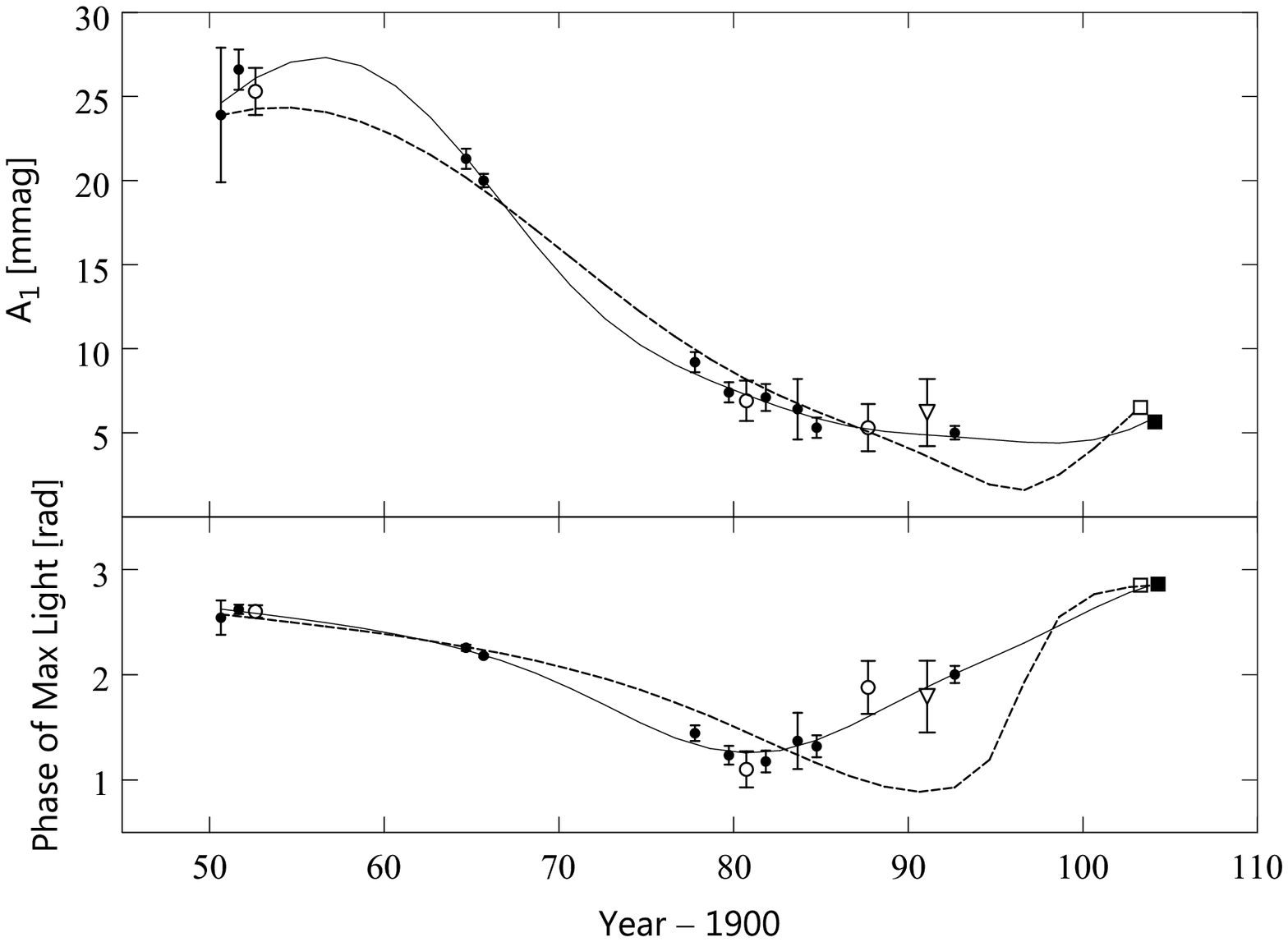} 
\caption{Upper panel: The amplitude of the $f_1$ term, $A_1$, computed from the parameters of the five-frequency fit of Table 4 
(solid line), compared with the yearly mean blue-light 1950-1992 $A_1$ (the dots, circles and the triangle) and the mean 2003 $y$ 
and $v$ $A_1$ (the filled and empty square, respectively). The dots, circles and the triangle are from figure 2 (top) of JP99 but 
with the error bars doubled. The 2003 error bars are not plotted because they would be of about the same size as the symbols. Also 
shown is the amplitude, computed using the parameters of a fit with the first three frequencies of Table 4 (dashed line). Lower 
panel: The same for the phase of maximum light of the $f_1$ term.  In both panels, the 2003 symbols are shifted along the abscissa 
by $\pm$0.5~yr to avoid overlap.}
\label{Fig12}
\end{figure}

\begin{figure} 
\includegraphics[width=80mm]{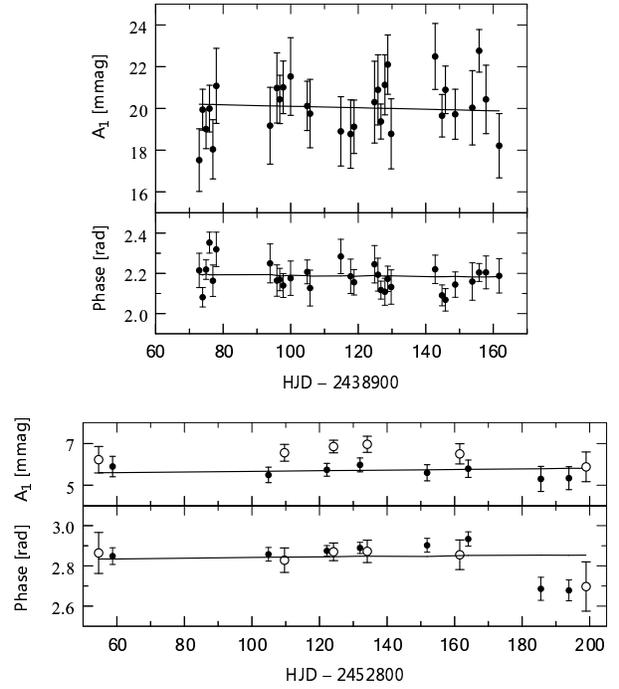} 
\caption{Upper half, upper panel: The amplitude of the $f_1$ term, $A_1$, computed from the parameters of the five-frequency fit of 
Table 4 (solid line), compared with the nightly 1965 $B$-filter $A_1$ from JP96 (dots). Lower half, upper panel: The same for the 
$y$ and $v$ $A_1$ derived in Section 3 (the dots and circles, respectively). Upper half, lower panel: The phase of maximum light of 
the $f_1$ term, computed from the parameters of the five-frequency fit of Table 4 (solid line), compared with the 1965 $B$-filter 
phases of maximum light (dots). Lower half, lower panel: The same for the $y$ and $v$ phases of maximum light (the dots and circles, 
respectively).}
\label{Fig13}
\end{figure}

The five-frequency fit accounts very well for the long-term variation of the amplitude and phase of the $f_1$ term. This can be seen 
from Fig.~\ref{Fig12} where the amplitude and the phase of maximum light computed from the parameters of Table 4 (solid lines) are 
compared with the yearly mean amplitudes and the yearly mean phases of maximum light (upper and lower panel, respectively). The 
observed and computed phases of maximum light, $\varphi_{\rm max}$, were obtained from the observed and computed epochs of maximum 
light, HJD$_{\rm max}$, using the formula
\begin{equation}
\varphi_{\rm max} = 2 \pi [E - (\rm{HJD}_{\rm max}-\rm{HJD}_0) $$f$$],
\end{equation}
where $E$ is the number of cycles which elapsed from an arbitrary initial epoch HJD$_0$ and $f = f_{1,1}$ from Table 4. The computed 
amplitude and phase of maximum light agree also with the nightly amplitudes and the nightly phases of maximum light. This is 
illustrated in the upper half of Fig.~\ref{Fig13} where the solid lines of Fig.~\ref{Fig12} are plotted together with the 1965 
$B$-filter amplitudes from JP96 and the 1965 $B$-filter phases of maximum light. The agreement between the computed  amplitude and 
the $y$ and $v$ amplitudes derived in Section 3 is also satisfactory (see the upper panel of the lower half of Fig.~\ref{Fig13}). 
The same goes for the computed and observed phases of maximum light (the lower panel of the lower half of the figure). Note that the 
lines in Figs.~\ref{Fig12} and \ref{Fig13} were not fitted to the points shown in the figures but were computed independently from 
the parameters of the five-frequency fit. From Fig.~\ref{Fig12} it is also clear that the three first frequencies of Table 4 alone 
would be insufficient to account for the variation of the phase, especially around 1900. This is an important conclusion because the 
first three periods in Table 4, $P_{1,1}$, $P_{1,2}$ and $P_{1,3}$, are very nearly equal to the periods $P_1$, $P^-_1$ and $P^+_1$ 
of L01, mentioned in the Introduction: the differences (in the sense `Table 4 {\em minus\/} L01') amount to 0.00000013, 0.00000011 
and 0.00000005 d, respectively. 

The first three frequencies of Table 4 form a very nearly equally-spaced triplet, $f_{1,2}$, $f_{1,1}$, $f_{1,3}$, with a mean 
spacing equal to 0.0137 yr$^{-1}$. The remaining frequencies, $f_{1,4}$ and $f_{1,5}$, flank the triplet at a distance of 0.0164 
yr$^{-1}$ from the first and the last frequency of the triplet, respectively. The triplet's spacing implies a time-scale of 73 yr 
for the long-term variation of the amplitude and phase of the $f_1$ term, while accounting for the non-sinusoidal shape of the 
variation seen in Fig.~\ref{Fig12} requires all five components. Note that the reciprocal of the time-span of the data is equal to 
1/53.4 = 0.0187 yr$^{-1}$, so that the spacings of the adjacent frequencies of the quintuplet amount to about 3/4 of the formal 
frequency resolution of the data. That they could be resolved nevertheless is due to their unequal amplitudes. 

\subsection{The $f_2$ term}

\begin{figure} 
\includegraphics[width=80mm]{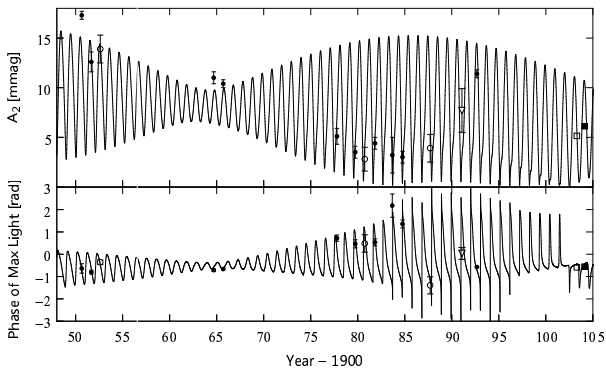} 
\caption{Upper panel: The amplitude of the $f_2$ term, $A_2$, computed from the parameters of the four-frequency fit of Table 5 
(solid line), compared with the yearly mean blue-light 1950-1992 $A_2$ (the dots, circles and the triangle) and the mean 2003 $y$ 
and $v$ $A_2$ (the filled and empty square, respectively). The dots, circles and the triangle are from figure 3 (top) of JP99 but 
with the error bars doubled. The 2003 error bars are not plotted because they would be of about the same size as the symbols. Lower 
panel: The same for the phase of maximum light of the $f_2$ term. In both panels, the 2003 symbols were shifted along the abscissa 
by $\pm$0.5~yr to avoid overlap.}
\label{Fig14}
\end{figure}

\begin{figure} 
\includegraphics[width=80mm]{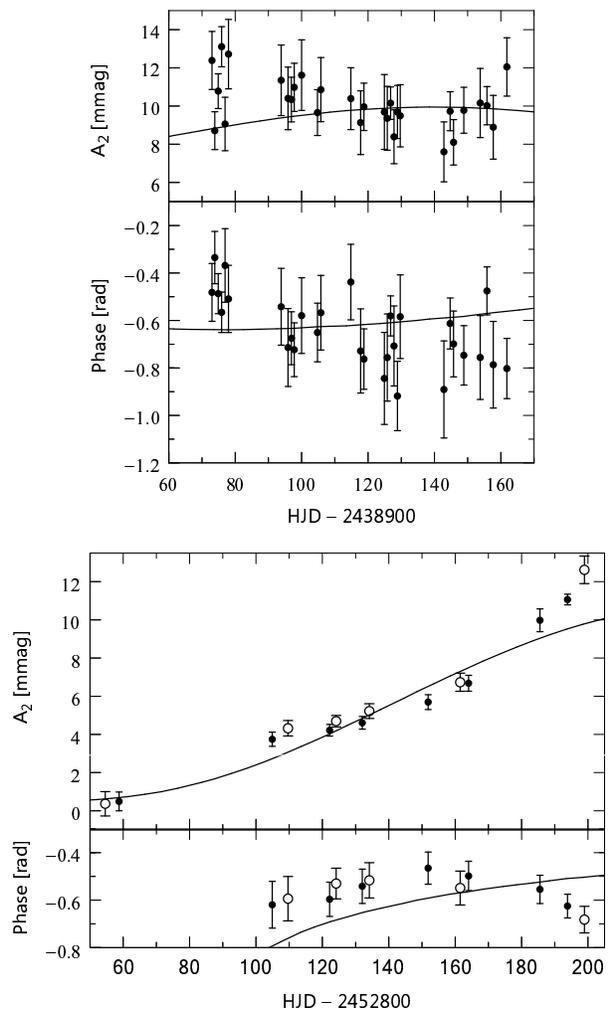} 
\caption{Upper half, upper panel: The amplitude of the $f_2$ term, $A_2$, computed from the parameters of the four-frequency fit of 
Table 5 (solid line), compared with the nightly 1965 $B$-filter $A_2$ from JP96 (dots). Lower half, upper panel: The same for the 
$y$ and $v$ $A_2$ derived in Section 3 (the dots and circles, respectively). Upper half, lower panel: The phase of maximum light of 
the $f_2$ term, computed from the parameters of the five-frequency fit of Table 5 (solid line), compared with the 1965 $B$-filter 
phases of maximum light (dots). Lower half, lower panel: The same for the $y$ and $v$ phases of maximum light (the dots and circles, 
respectively). The first-segment $y$ and $v$ phases of maximum light, equal to $-$1.6$\,\pm\,$1.0 and 1.7$\,\pm\,$1.7 rad, 
respectively, are not shown.}
\label{Fig15}
\end{figure}

Frequency analysis of $B\&y -{}$13 yielded four frequencies before the noise prevented detecting further frequencies. However, a 
four-frequency fit did not account very well for the variation of the amplitude and phase of the $f_2$ term. After a number of 
trials, we found that the fit improved when the amplitudes were assumed to vary uniformly with time, i.e.\ when constant amplitudes 
in the observational equations were replaced by $A_{2,j}+B_{2,j} t$. The parameters of a least-squares fit with the amplitudes 
modified in this way are listed in Table 5. The standard deviation of the fit amounted to 5.6 mmag. The epochs of observations were 
reckoned from HJD\,244\,5784. A comparison of the amplitudes and phases of maximum light computed from the parameters of Table 5 
(solid lines) with the observed amplitudes and phases of maximum light is shown in Figs.~\ref{Fig14} and \ref{Fig15}. The 
observed and computed phases of maximum light were obtained from the observed and computed maxima using equation (5) with $f = 
f_{2,1}$ from Table 5. The agreement between the computed and observed amplitudes and phases of maximum light of the $f_2$ term seen 
in Figs.~\ref{Fig14} and \ref{Fig15} is less satisfactory than was the case for $f_1$ (Figs.~\ref{Fig12} and \ref{Fig13}). 

As can be seen from Table 5, $f_{2,1}-f_{2,2} ={}$0.002607 and $f_{2,4}-f_{2,1} ={}$0.002648 d$^{-1}$, so that $f_{2,2}$, $f_{2,1}$ 
and $f_{2,4}$ form a very nearly equidistant frequency triplet, with a mean separation equal to 0.002628 d$^{-1}$. Although this 
number is close to 1 yr$^{-1}$, it is not an artefact because the aliases were removed in our procedure of pre-whitening. The beat 
period corresponding to the mean separation of the triplet is equal to 380.5 d, a value about 10\,\% greater than the beat-period 
between the L01 periods $P_2$ and $P^+_2$. The difference between $f_{2,2}$ and $f_{2,3}$ is equal to 0.000063 d$^{-1}$ or 0.023 
yr$^{-1}$, implying a time scale of 43 yr. The latter number is close to the time scale of the variation of $A_2$ derived by JP99 
from the 1950-1992 data. 

\begin{table*}
 \centering
 \begin{minipage}{130mm}
\caption{Fine structure of the $f_2$ term.}
\begin{center}
  \begin{tabular}
{rccr@{\hspace{3pt}$\pm$\hspace{3pt}}lr@{\hspace{3pt}$\pm$\hspace{3pt}}lr@{\hspace{3pt}$\pm$\hspace{3pt}}l} \\
  \hline
  $j$ & \multicolumn{1}{c}{$f_{2,j}$ [d$^{-1}$]}& \multicolumn{1}{c}{$P_{2,j}$ [d]} & \multicolumn{2}{c}
{$A_{2,j}$ [mmag]} & \multicolumn{2}{c}{$B_{2,j}$ [mmag/d]} & \multicolumn{2}{c}{$\Phi_{2,j}$ [rad]} \\
 \hline
  1&\hspace{1pt} 5.855574&0.1707775&6.828 &0.093&$-$0.000173&0.000021&5.017&0.013\\
  2&\hspace{1pt} 5.852967&0.1708535&3.592 &0.096&   0.000047&0.000016&5.199&0.028\\
  3&\hspace{1pt} 5.852904&0.1708554&3.008 &0.090&$-$0.000074&0.000023&5.483&0.028\\
  4&\hspace{1pt} 5.858222&0.1707003&2.164 &0.096&   0.000108&0.000016&3.836&0.043\\
\hline
\end{tabular}
\end{center}
\end{minipage}
\end{table*}

\subsection{The $f_3$ term}

\begin{figure} 
\includegraphics[width=80mm]{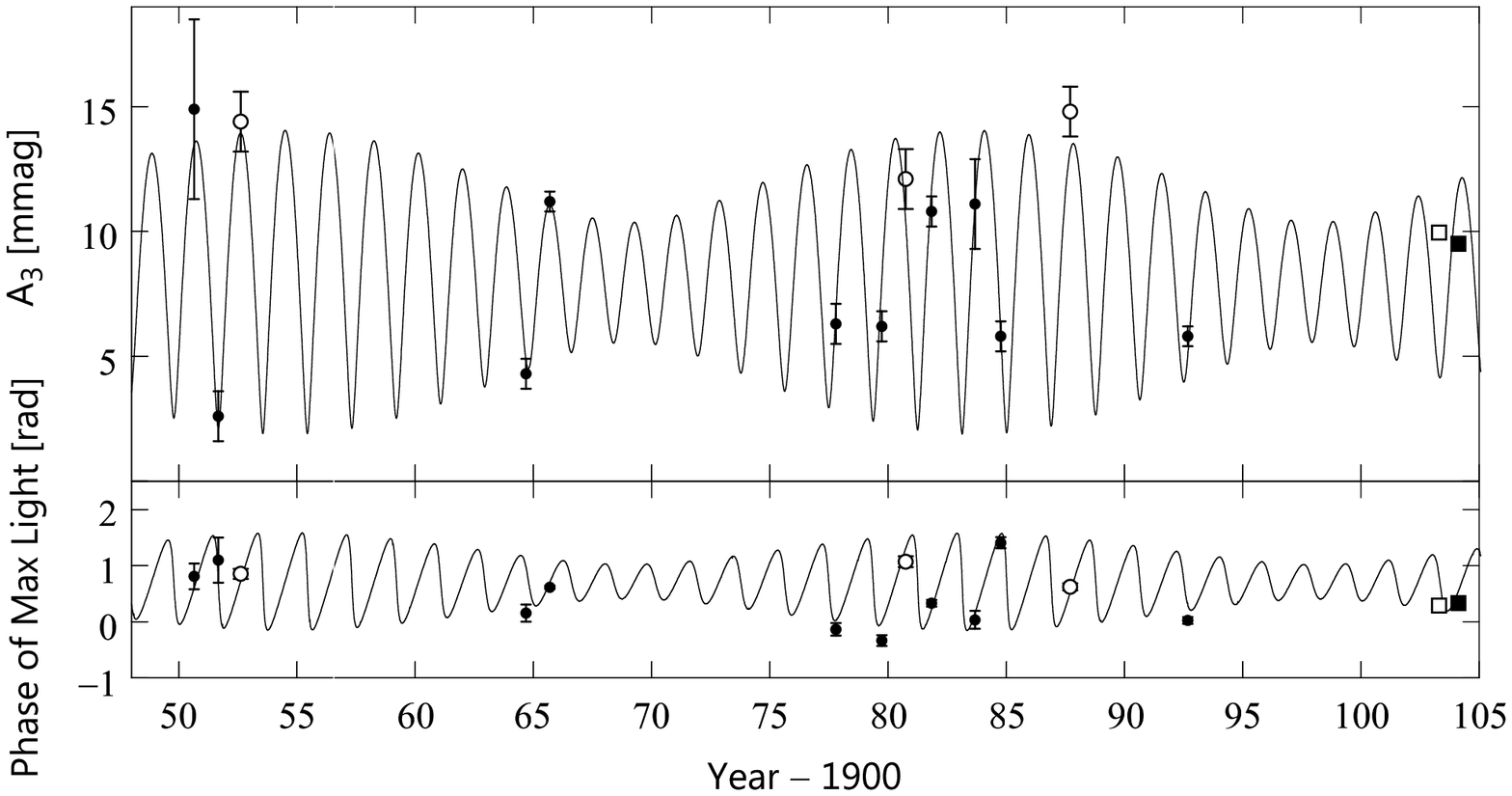} 
\caption{Upper panel: The amplitude of the $f_3$ term, $A_3$, computed from the parameters of the three-frequency fit of Table 5 
(solid line), compared with the yearly mean blue-light 1950-1992 $A_3$ from table A1 of JP96 and table 1 of JP99 (the dots and 
circles; the latter represent data of lower weight) and the mean 2003 $y$ and $v$ $A_3$ (the filled and empty square, respectively). 
The 2003 error bars are not plotted because they would be of about the same size as the symbols. Lower panel: The same for the phase 
of maximum light of the $f_3$ term. In both panels, the 2003 symbols were shifted along the abscissa by $\pm$0.5 yr to avoid 
overlap.}
\label{Fig16}
\end{figure}

\begin{figure} 
\includegraphics[width=80mm]{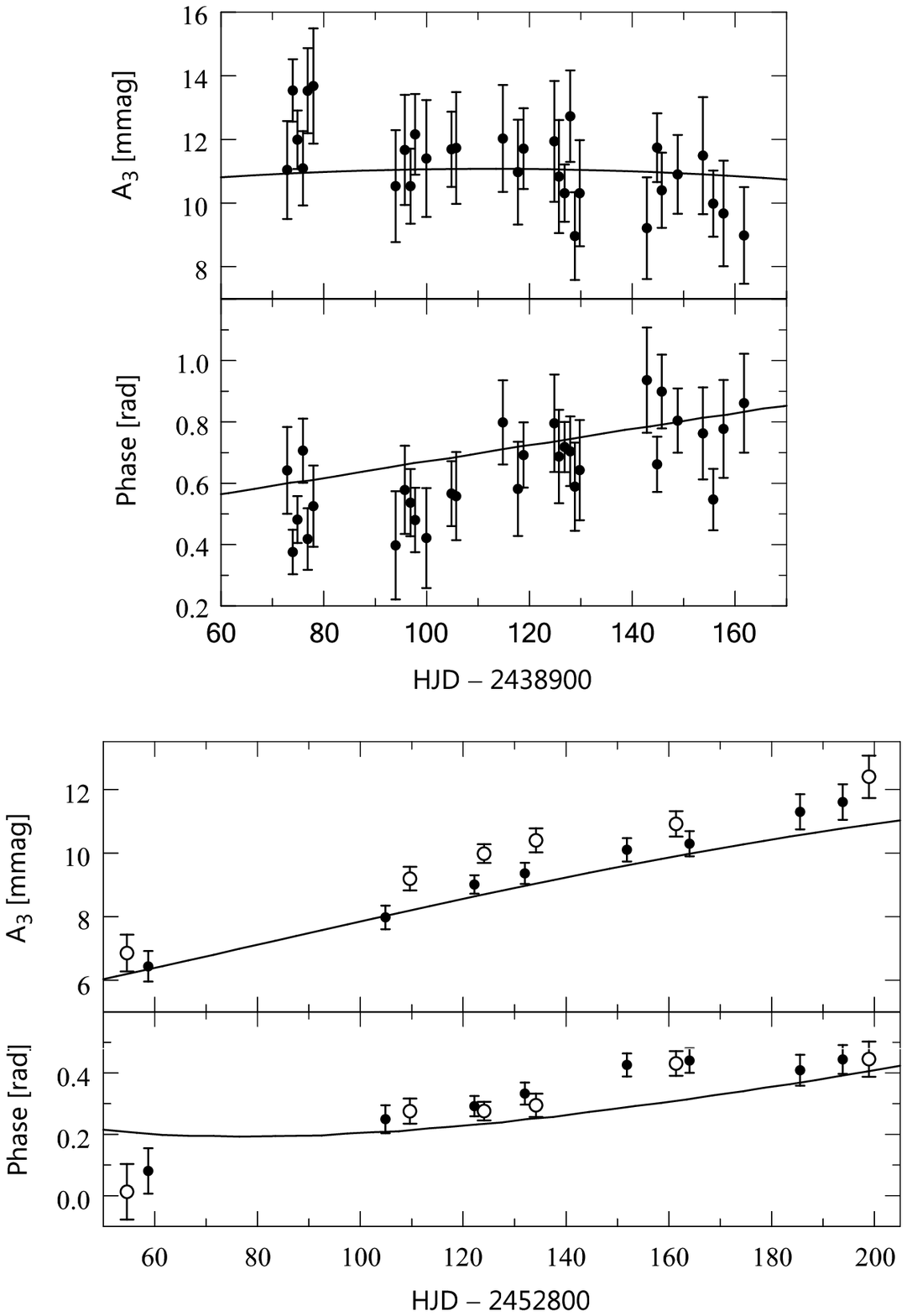} 
\caption{Upper half, upper panel: The amplitude of the $f_3$ term, $A_3$, computed from the parameters of the three-frequency fit of 
Table 5 (solid line), compared with the nightly 1965 $B$-filter $A_3$ from JP96 (dots). Lower half, upper panel: The same for the 
$y$ and $v$ $A_3$ derived in Section 3 (the dots and circles, respectively). Upper half, lower panel: The phase of maximum light of 
the $f_3$ term, computed from the parameters of the three-frequency fit of Table 5 (solid line), compared with the 1965 $B$-filter 
phases of maximum light (dots). Lower half, lower panel: The same for the $y$ and $v$ phases of maximum light (the dots and circles, 
respectively).}
\label{Fig17}
\end{figure}

As we mentioned in the Introduction, the $f_3$ term was found by JP96 to be a doublet. Using all blue-filter observations of 16 Lac 
available at the time, JP99 determined the doublet frequencies to be $f_{3,1} ={}$5.5025779$\,\pm\,$0.0000005 and $f_{3,2} 
={}$5.5040531$\,\pm\,$0.0000008 d$^{-1}$. Slightly different frequencies, equal to 5.5025928 and 5.5040765 d$^{-1}$, were 
subsequently derived by L01 from RV data. A frequency analysis of $B\&y -{}$12 showed this term to be a triplet. Using the 
frequencies read off the amplitude spectra as starting values in a three-frequency nonlinear least-squares fit of equation (1) to 
$B\&y -{}$12 resulted in the frequencies, amplitudes and phases listed in Table 5. The standard deviation of the fit amounted to 5.4 
mmag. The epochs of observations were reckoned from HJD\,244\,5784. A comparison of the amplitudes and phases of maximum light 
computed from the parameters listed in Table 5 with observations is shown in Figs.~\ref{Fig16} and \ref{Fig17}. The observed 
and computed phases of maximum light were obtained from the observed and computed maxima using equation (5) with $f = f_{3,1}$ from 
Table 5. 

\begin{table*}
 \centering
 \begin{minipage}{130mm}
\caption{Fine structure of the $f_3$ term.}
\begin{center}
  \begin{tabular}
{rr@{\hspace{3pt}$\pm$\hspace{3pt}}lr@{\hspace{3pt}$\pm$\hspace{3pt}}lr@{\hspace{3pt}$\pm$\hspace{3pt}}l
r@{\hspace{3pt}$\pm$\hspace{3pt}}l} \\
  \hline
  $j$ & \multicolumn{2}{c}{$f_{3,j}$ [d$^{-1}$]}& \multicolumn{2}{c}{$P_{3,j}$ [d]} & \multicolumn{2}{c}
{$A_{3,j}$ [mmag]} & \multicolumn{2}{c}{$\Phi_{3,j}$ [rad]} \\
 \hline
  1&\hspace{1pt} 5.50257795&0.00000033&0.181733000& 0.000000011&7.97&0.10&1.914&0.013\\
  2&\hspace{1pt} 5.50405957&0.00000058&0.181684080& 0.000000019&4.24&0.10&2.484&0.024\\
  3&\hspace{1pt} 5.50396544&0.00000127&0.181687187& 0.000000042&1.84&0.09&2.348&0.049\\
\hline
\end{tabular}
\end{center}
\end{minipage}
\end{table*}

The new values of $f_{3,1}$ and $f_{3,2}$ are very nearly equal to those obtained by JP99. The new value of the beat-period is equal 
to 674.94$\,\pm\,$0.30 d, in good agreement with the beat-period between the L01's periods $P_3$ and $P^+_3$, mentioned in the 
Introduction. The third frequency, placed asymmetrically between $f_{3,1}$ and $f_{3,2}$, gives rise to two beat-periods, 
720.73$\,\pm\,$0.35 d and 29.09$\,\pm\,$0.43 yr.

\section{Summary and discussion}

Over the 179.2-d interval spanned by the present multisite $uvy$ data, the amplitude of the $f_1$ term was constant but the 
amplitudes of the $f_2$ and $f_3$ terms, $A_2$ and $A_3$, increased by several mmags (see Figs.~\ref{Fig02} and \ref{Fig03}). The 
latter fact made the usual procedure of pre-whitening inapplicable. In Section 3, using the values of the three frequencies 
determined from all available $B$, $b$ and $y$ data (Section 7), we pre-whitened the data piecewise, in segments so long that the 
three terms could be resolved but short enough to neglect the variation of $A_2$ and $A_3$. In the pre-whitened data we detected two 
low-amplitude terms having frequencies $f_4 ={}$11.3582 and $f_5 ={}$6.2990 d$^{-1}$. The former frequency is equal to $f_2+f_3$, 
the latter is new. The amplitudes of the $f_4$ term amount to 0.57$\,\pm\,$0.09, 0.64$\,\pm\,$0.10 and 0.92$\,\pm\,$0.15 mmag for  
$y$, $v$ and  $u$, respectively, while those of $f_5$, to 0.64$\,\pm\,$0.09, 0.61$\,\pm\,$0.10 and 1.05$\,\pm\,$0.15 mmag, 
respectively. The frequencies 2$f_{\rm orb}$, 2$f_1$, $f_1+f_3$ and $f_1+f_2$ seen in 1965 (see the Introduction) were not found. 
Why $f_2+f_3$ was present in 1965 and 2003-2004 while 2$f_1$, $f_1+f_3$ and $f_1+f_2$ were absent in 2003-2004 is easy to 
understand: in 2003-2004 $A_2$ was moderately smaller and $A_3$ was slightly smaller than in 1965 (see Figs.~\ref{Fig14} and 
\ref{Fig16}) while $A_1$ decreased between 1965 and 2003-2004 to about one fourth of its 1965 value (see Fig.~\ref{Fig12}). The fact 
that 2$f_{\rm orb}$ was missing in 2003-2004 suggests that it may be related to $f_1$. Finally, $f_5$ was not detected in 1965 
because its $V$ and $B$ amplitudes were then below $\sim$0.40 mmag.

With the $f_i$ $(i = 1,..,5)$ terms taken into account, we repeated the piecewise pre-whitening. The resulting residuals were used 
in Section 4 to plot the eclipse light-curves (Figs.~\ref{Fig05} and \ref{Fig06}). The light curves show no ellipticity effect and 
no secondary eclipse can be detected. However, a marginal reflection effect is present. From the $y$ light-curves and the L01 
spectroscopic elements $\omega$ and $e$, we computed the relative radius of the primary, $r_{\rm p}$, the orbital inclination, $i$, 
and the reflected light from the secondary, $S_{\rm s}$, for a range of $k = r_{\rm s}/r_{\rm p}$ by means of {\sc EBOP} 
\citep{etz}. This allowed an examination of the position of the secondary component in the HR diagram in relation to pre-MS 
evolutionary tracks (Fig.~\ref{Fig07}), leading to the conclusions that (1) the secondary component is in the pre-MS phase of its 
evolution, and (2) the parameters of the system can be constrained to 0.21${}\la k \la{}$0.29, 0.125${}\la r_{\rm p} \la{}$0.132 and 
83\fdg4${}\ga i \ga{}$82\fdg0.  

For $f_1$, $f_2$ and $f_3$, the $uvy$ and the Geneva $UBG$ amplitude ratios are derived from the multisite data and compared with 
the theoretical ones for the spherical-harmonic degree $\ell ={}$0,..,4 in Section 6. The theoretical amplitude ratios were computed 
using $T_{\rm eff}$ and $\log g$ of 16 Lac obtained in Section 5. In Section 6, the highest degree, $\ell ={}$4, is shown to be 
incompatible with the observations. The first term, $f_1$, could be identified with an $\ell ={}$0 mode, while the third, $f_3$, 
with an $\ell ={}$1 mode. In the case of $f_2$, an unambiguous spherical-harmonic degree identification was not possible: it can be 
either an $\ell ={}$2 or 3 mode, with the latter possibility less likely because of the effect of cancellation in integrating over 
the stellar disc. 

In Section 7, using the present $y$-filter magnitudes and archival blue-filter magnitudes, we investigate the long-term variation of 
the amplitudes and phases of the three high-amplitude terms over the interval of 53.4 yr spanned by the data. In the case of $f_1$, 
the magnitudes can be represented by means of a sum of five sinusoidal components with closely spaced frequencies (see Table 4). The 
first three frequencies form an equally-spaced triplet with a spacing of 0.0137 yr$^{-1}$, implying a time-scale of 73 yr, in 
agreement with JP99 and L01. The sum of the five components accounts very well for the non-sinusoidal shape of the variation seen in 
Fig.~\ref{Fig12}. Since the $f_1$ mode is radial, the regularities in the frequency spacings suggest an underlying amplitude and 
phase modulation of a single pulsation mode. 

In the case of $f_2$ we could represent the 1950-2003 magnitudes by a sum of four sinusoidal components with uniformly variable 
amplitudes (see Table 5). Of these, three components ($f_{2,2}$, $f_{2,1}$ and $f_{2,4}$ in the order of increasing frequency) form 
a triplet very nearly equidistant in frequency with a mean separation of 0.002628 d$^{-1}$, corresponding to a beat period of 380.5 
d, while the $f_{2,3}$ component precedes the first member of the triplet by 0.023 yr$^{-1}$, corresponding to a beat-period of 43 
yr. The 380.5-d beat-period dominates the variation of the amplitude and phase (see Fig.~\ref{Fig14}); the 43-yr beat-period is 
close to the time scale of the variation of $A_2$ derived by JP99 from the 1950-1992 data.  The $f_{2,2}$, $f_{2,1}$, $f_{2,4}$ 
triplet may be the result of (1) rotational splitting of a nonradial mode, (2) an accidental coincidence of two nonradial modes, one 
rotationally split, and (3) an accidental coincidence of three nonradial modes. In all cases, only some members of the rotationally 
split multiplets would be excited to observable amplitudes. In (1) and (2), the velocity of the star's rotation can be computed from 
the Ledoux first-order formula using the radius from the eclipse solutions. The results are $V_{\rm rot} \la{}$1 km s$^{-1}$ if (1), 
and $V_{\rm rot} \la{}$2 km s$^{-1}$ if (2). These numbers are in severe conflict with the observed $V_{\rm rot} \sin i$  
\citep{GGS}. Thus, we are left with (3) and the problem why three accidentally coincident modes should be nearly equidistant 
in frequency. Finally, there is the possibility that our representation by means of a sum of four sinusoidal components may be 
merely a formal description of the complex long-term variation of the amplitude and phase of the $f_2$ term.

JP96 have noted that the reciprocal of the growth rates of several $\ell \le 2$ pulsation modes which are excited in models of 16 
Lac and can be identified with the $f_1$ and $f_2$ terms are of the same order of magnitude as the time scales of 73 and 43 yr just 
mentioned. Since amplitude and phase modulation on a time-scale of the order of the reciprocal of the growth rate is predicted by 
the theory of non-linear interaction of pulsation modes, JP96 suggested that these time scales in the observed long-term behaviour 
of the $f_1$ and $f_2$ modes result from (1) the 1:1 resonance between them, or (2) a resonant coupling to other modes. These points 
still await theoretical verification. 

Finally, the $f_3$ 1950-2003 magnitudes could be represented by a sum of three sinusoidal components (see Table 5). The first two 
components have frequencies $f_{3,1}$ and $f_{3,2}$ very nearly equal to those derived by JP96 and JP99 from the data available at 
the time. These two frequencies produce a beat period of 720.73$\,\pm\,$0.35 d which dominates the amplitude and phase variation 
(see Fig.~\ref{Fig16}). The frequencies $f_{3,3}$ and $f_{3,2}$ produce a beat period of 29.09$\,\pm\,$0.43 yr. As in the case 
of $f_2$, assuming that the $f_{3,1}$, $f_{3,2}$ doublet is due to rotational splitting of a nonradial mode leads to $V_{\rm rot} 
\la{}$1 km s$^{-1}$, much smaller than observed. JP96 suggested that the doublet represents an accidental near-coincidence of two 
self-excited modes. In view of the unambiguous identification of the harmonic degree of the $f_3$ term (Section 6), these two modes 
must both have $\ell ={}$1. On the other hand, the members of the $f_{2,2}$, $f_{2,1}$, $f_{2,4}$ triplet may have different $\ell$, 
a situation encountered earlier in 1 Mon \citep{BS80} and 12 Lac \citep{HJR06}. 

\section*{Acknowledgments}

We are indebted to Dr.\ Paul B.\ Etzel for providing the source code of his computer program {\sc EBOP} and explanations. GH was 
supported by the Polish NCN grant 2011/01/B/ST9/05448. AP, ZK and GM acknowledge the support from the NCN grant 2011/03/B/ST9/02667. 
In this research, we have used the Aladin service, operated at CDS, Strasbourg, France, and the SAO/NASA Astrophysics Data System 
Abstract Service.

\appendix
\section{The variable comparison star 2 Andromedae}

\subsection{Frequency analysis}

\begin{figure}
\includegraphics[angle=0,width=88mm,viewport=-00 5 250 250]{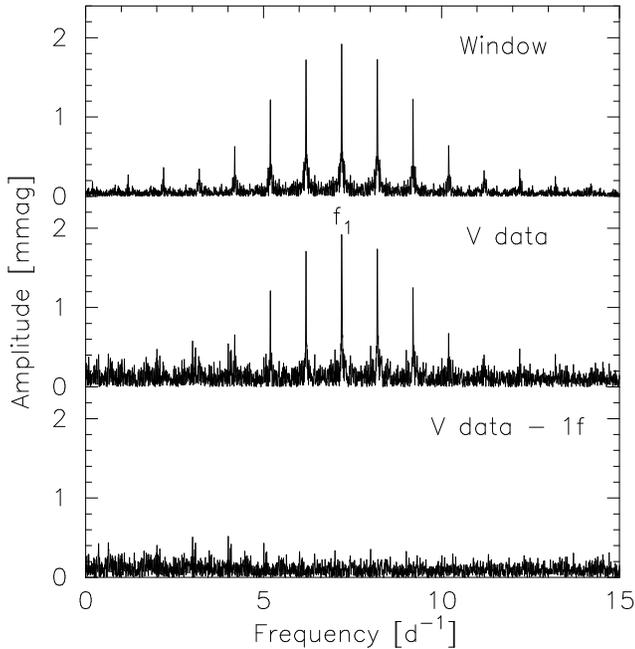}
\caption[]{Amplitude spectrum of 2 And. The top panel shows the spectral window of the data, followed by the periodogram of the data 
in the middle panel. Pre-whitening with $f_1$ leaves only noise in the residuals (bottom panel), aside from some residual extinction 
effects (peaks at 2.0, 3.0, 4.0 and 5.0 d$^{-1}$).}
\label{FigA1}
\end{figure}

As mentioned in the Introduction, the comparison star 2~And turned out to be a small-amplitude variable. Fig.~\ref{FigA1} 
shows the results of a frequency analysis of the differential $y$ magnitudes 2~And $-$ 10 Lac. The less accurate Sierra Nevada and 
Piszk\'estet\H o observations were omitted.

The variability of 2 And can be described by a single frequency (Fig.~\ref{FigA1}), which is confirmed by the analysis of the 
measurements in the other filters. Nonlinear least-squares fitting to our $y$-filter data gives a value of 
7.195301$\,\pm\,$0.00012~d$^{-1}$, whereby the error estimate was computed with the formulae of \citet{MOD99}. Fitting a sine curve 
of this frequency to the data results in the amplitudes and relative phases, $\phi_x - \phi_y$ ($x ={}u$, $v$, $U$, $B1$, $B$, $B2$, 
$V1$, $V$, $G$), listed in Table A1.

\begin{table}

\caption[]{Amplitudes and relative phases of the single frequency 7.195301 d$^{-1}$ for the variable comparison 
star 2 And. Formal error estimates \citep[following][]{MOD99} for the amplitudes and phases are given.}

\begin{center}
\begin{tabular}{lcc}
\hline
 Filter & Amplitude & $\phi_x - \phi_y$ \\
 & [mmag] & [rad] \\
\hline
$u$ &  1.72$\,\pm\,$0.13 &$-$0.16$\,\pm\,$0.09 \\
$v$ &  1.13$\,\pm\,$0.08 &$-$0.05$\,\pm\,$0.08 \\
$y$ &  1.91$\,\pm\,$0.07 &\hspace{5pt}0 by definition \\
$U$ &  2.07$\,\pm\,$0.29 &\hspace{7pt}0.07$\,\pm\,$0.14 \\
$B1$ & 1.57$\,\pm\,$0.25 &$-$0.17$\,\pm\,$0.16 \\
$B$ &  1.69$\,\pm\,$0.25 &$-$0.06$\,\pm\,$0.15 \\
$B2$ & 1.65$\,\pm\,$0.26 &$-$0.22$\,\pm\,$0.16 \\
$V1$ & 2.26$\,\pm\,$0.25 &$-$0.02$\,\pm\,$0.12 \\
$V$ &  2.05$\,\pm\,$0.22 &\hspace{7pt}0.01$\,\pm\,$0.12 \\
$G$ &  2.42$\,\pm\,$0.26 &$-$0.04$\,\pm\,$0.11 \\
\hline
\end{tabular}
\end{center}
\end{table}

\subsection{The cause of variability of 2 And}

Given the spectral type of A3\,Vn and the period of the variability, one would immediately suspect that the light variations of 2 
And are due to $\delta$~Scuti-type pulsation. In fact, the star was so classified by \citet{HJR06}. However, if 2~And were a 
$\delta$~Scuti variable, the amplitudes are generally expected to increase from $y$ to $v$ and from $G$ to $B1$, and then 
level out or drop again towards $u$ and $U$ \citep[e.g., see][]{H94}. Interestingly, just the inverse is the case. In the 
following we investigate possible causes for this observation.

2 And is not a single star. It is a close visual binary discovered over a century ago \citep{B1894}. The two components are 
physically associated. The latest determination of the orbital period is 74 yr with an eccentricity of 0.8 and a semimajor axis of 
0.23\arcsec \citep{RR10}. Transforming the Tycho-2 photometry of the two components \citep{FM00}, to the standard Johnson 
system according to \citet{Be00} gives: $V_{\rm A} ={}$5.24 mag, $(B-V)_{\rm A} ={}$0.07 mag for 2 And A, and $V_{\rm B} ={}$7.51 
mag, $(B-V)_{\rm B} ={}$0.23 mag for 2 And B. Therefore $V_{\rm A+B} ={}$5.113 mag and $(B-V)_{\rm A+B} ={}$0.086 mag, in reasonable 
agreement with the measured $V ={}$5.100 mag and $B-V ={}$0.094 mag for the system \citep{m91}.

The revised {\em Hipparcos} parallax of 2 And \citep{vL} implies a distance of 129$\,\pm\,$9~pc. Concerning reddening, a comparison 
of the results from the galactic reddening law of \citet{Ch} for 2 And, and the reddening of two stars within 4$^{\rm o}$ of 2 And 
in the sky, at a similar {\em Hipparcos} distance (HR\,8870 and HD\,218394) lead us to adopt $E(B-V) ={}$0.022$\,\pm\,$0.022 mag and 
$A(V) ={}$0.07$\,\pm\,$0.07 mag.

\begin{figure}
\includegraphics[width=84mm,viewport=00 05 262 267]{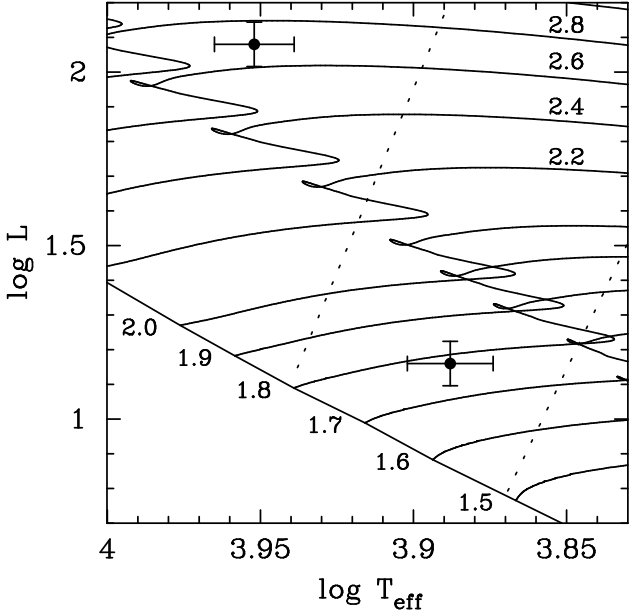}
\caption[]{The positions of 2 And A and B in the theoretical HR diagram.  Some stellar model evolutionary tracks labeled with their 
masses in solar units (solid lines) are included. The borders of the $\delta$ Scuti instability strip \citep[adopted from][dotted 
lines]{R+00} are included for comparison. The solid diagonal line is the ZAMS.}
\label{FigA2}
\end{figure}

We then arrive at $M_{\rm V} ={}-$0.39$\,\pm\,$0.16 mag for 2 And~A and 1.88$\,\pm\,$0.16 mag for 2 And B. The relations of 
\citet{Fl96} yield estimates of $T_{\rm eff} ={}$8950$\,\pm\,$250~K and $M_{\rm bol} ={}-$0.45$\,\pm\,$0.16 mag for 2 And A, and 
7720$\,\pm\,$250~K and 1.85$\,\pm\,$0.16 mag for 2 And B. The positions of the two components are shown in a theoretical HR diagram 
in Fig.~\ref{FigA2}. The evolutionary tracks plotted in this figure were computed with the Warsaw-New Jersey stellar evolution 
code, the OPAL equation of state and the OPAL opacity tables \citep{IR96}, a hydrogen abundance of $X ={}$0.7, a metal abundance of 
$Z ={}$0.012 \citep{A+04}, and no convective core overshooting. We assumed $V_{\rm rot} ={}$ 250~km~s$^{-1}$ on the zero-age main 
sequence, so that the observed $V_{\rm rot} \sin i ={}$212 km~s$^{-1}$ \citep{RZG07} would be approximately matched.

One finds that 2 And A is an $M ={}$2.7$\,\pm\,$0.1~M$_{\sun}$ star that probably has already left the main sequence ($\log g 
={}$3.40$\,\pm\,$0.12, $\log {\rm age} ={}$8.57$\,\pm\,$0.04). Including core overshooting would decrease $M$ and the age but $\log 
g$ would remain virtually unchanged. 2 And A is located off the $\delta$~Scuti instability strip. 2 And B is situated in the centre 
of the strip ($M ={}$1.78$\,\pm\,$0.06~M$_{\sun}$, $\log g ={}$3.90$\,\pm\,$0.16, $\log {\rm age} ={}$8.84$^{+0.07}_{-0.18}$).

With these parameters in hand, we shall test four possible hypotheses to explain the cause of the variability of 2 And: 
$\delta$~Scuti pulsation or ellipsoidal variability, of either 2 And A or 2 And B. Rotational modulation can be excluded immediately 
because either star would have to rotate faster than its breakup velocity.

Regarding the hypothesis of $\delta$~Scuti pulsation, we computed theoretical pulsation amplitudes of $\delta$~Scuti models for the 
Str\"omgren and Geneva passbands in the parameter ranges depicted in Fig.~\ref{FigA2} for an assumed radial velocity amplitude, 
using the methodology by \citet{BE99}. We note that whilst computing realistic absolute amplitudes is still out of reach, amplitude 
{\it ratios} are easily obtainable.

We computed theoretical pulsation amplitudes of $\delta$~Scuti models for the Str\"omgren and Geneva passbands. From the flux ratio 
of 2 And A to 2 And B in the $y$ band (estimated from $V_{\rm A}$ and $V_{\rm B}$) and the observed $y$ amplitude from Table A1 we 
then determined the undiluted photometric amplitudes of 2 And A and 2 And B in the Str\"omgren $y$ band, ($A_{\rm true} ={}$ 0.0022 
and 0.0215 mag, respectively), and used the model amplitudes to predict the intrinsic amplitudes in the other bands. As the next 
step, we determined the flux ratio of the two components of 2 And in the different passbands. To this end, we used \citet{Ku94} 
model atmospheres, representing 2 And A with a $T_{\rm eff} ={}$9000~K, $\log g ={}$3.5 model atmosphere, and 2 And B with a $T_{\rm 
eff} ={}$7750~K, $\log g ={}$4.0 model atmosphere. We then integrated the monochromatic fluxes from these model atmospheres over the 
ten photometric passbands used, and scaled the resulting fluxes to the observed ratio in $y$.

The results of these computations can be summarized as follows: the behaviour of the pulsation amplitudes with wavelength is 
inconsistent with the observations for any kind of assumed pulsation of 2 And A. The reason is that the temperature of the star is 
so high, that the pulsation amplitudes always increase towards shorter wavelength, no matter which spherical degree of the pulsation 
was tested (we checked for $0\leq l \leq 8$). The additional flux of 2 And B does not change this behaviour due to the considerable 
luminosity difference.

Assuming that 2 And B was a $\delta$ Scuti pulsator, and considering the amplitude contamination due to the light of 2 And A, we 
find that the observed pulsation amplitude vs. wavelength dependence is only consistent with modes of even spherical degree of $l 
\geq 6$. We recall that in such a scenario the intrinsic pulsation amplitude of 2 And B in the Str\"omgren $y$ band, $A_{\rm true}$ 
is 0.0215 mag. Geometric cancellation of modes with $l \geq 6$ decreases the observed amplitudes to less than 1/50 of the intrinsic 
value \citep{DD+02}, i.e. the intrinsic pulsation amplitude would be enormous in this case. We therefore consider $\delta$ Scuti 
pulsation of either of the two A-type stars in the 2 And system as highly unlikely.

A remaining possibility is an ellipsoidal variation of either component. Consequently, we repeated the previous procedure under the 
assumption of no colour dependence of the amplitude. We found that the resulting wavelength-amplitude dependence explains the 
observations much better than the $\delta$ Scuti-pulsation hypothesis. This implies an orbital period of twice the observed value. 
Using the effective temperatures, the luminosities and the masses derived above we find from the Kepler's third law that in the case 
of 2 And A the sum of the radii of the components Aa and Ab, $R_{\rm Aa} + R_{\rm Ab}$, would be a factor of about 2 greater than 
the semimajor axis of the relative orbit, $a$, regardless of whether we assume (1) equal luminosity and mass components, or (2) the 
secondary component much fainter and less massive than the primary component. In the case of 2 And B, assumption (1) leads to 
$R_{\rm Ba} + R_{\rm Bb} \approx{}$ 1.5$a$, while assuming (2) one gets $R_{\rm Ba} \approx{}$1.0$a$. Thus, an ellipsoidal variation 
of 2 And B, arising from a tidal distortion of component Ba by a much less massive secondary component Bb in a tight orbit may be 
the cause of the variability of 2 And.

In order to examine the last possibility let us use the lowest order approximation, certainly adequate in the present case, 
according to which the amplitude of the light variation due to aspect changes of the tidally distorted primary can be expressed by 
means of the following formula:
\begin{equation} 
\delta m = 1.629 A_\lambda q (R_{\rm Ba}/a)^3 \sin^2 i, 
\end{equation}  
where $\delta m$ is expressed in magnitudes, $A_\lambda$ is the photometric distortion parameter of \citet{RM52}, $q$ is the mass 
ratio of the components, and $i$ is the inclination of the orbit to the tangent plane of the sky. We note that this formula is 
equivalent to equation (6) of \citet{Mo85}. Neglecting the light variation of the secondary, we have $\delta m = A_{\rm true}$. 
Then, taking into account the fact that in the present case $A_\lambda (R_{\rm Ba}/a)^3 \approx{}$1, we get $q \approx {}$0.013 and 
0.052 for $i ={}$90\degr and 30\degr, respectively. The corresponding masses of the secondary component are equal to $\sim$0.03 and 
$\sim$0.11 M$_{\sun}$, i.e.\ they are in the range of the masses of brown dwarfs. Thus, our hypothesis that 2 And B is an 
ellipsoidal variable leads to the following model: the 2 And Bb component is a brown dwarf in a tight orbit around 2 And Ba, a late 
A or an early F star. The amplitude of the RV variation, $K_1 \sin i$, would be equal to 10 and 20 km~s$^{-1}$ for $i ={}$90\degr 
and 30\degr, respectively. According to the ephemeris provided by \citet{RR10}, the separation of the components A and B will be 
0.135\arcsec in 2014 January, decreasing to 0.046\arcsec by 2018. In the near future, it will be thus impossible to get a 
spectrogram of 2 And B without an overwhelming contribution from 2 And A. Consequently, measuring the RV of 2 And B may be very 
difficult.

\bsp

\label{lastpage}

\end{document}